\documentclass[12pt]{article}
\usepackage{amssymb,amsmath}
\usepackage{comment}
\usepackage{graphicx}
\usepackage{bm}
\usepackage{color}
\usepackage{enumerate}
\usepackage{subfigure}
\usepackage{hyperref}

\numberwithin{equation}{section}

\begin{document}

\thispagestyle{empty}
\begin{flushright}
DCPT-15/69 \\
NCTS-TH/1508
\\

\end{flushright}
\vskip2cm
\begin{center}
{\Large Topological M-Strings and Supergroup WZW Models}

\vskip1.5cm
Tadashi Okazaki\footnote{tadashiokazaki@phys.ntu.edu.tw} 

\bigskip
{\it 
Department of Physics and Center for Theoretical Sciences,\\
National Taiwan University, Taipei 10617, Taiwan}
\\
\bigskip
and
\\
\bigskip
Douglas J. Smith\footnote{douglas.smith@durham.ac.uk}

\bigskip
{\it Department of Mathematical Sciences, Durham University,\\
Lower Mountjoy, Stockton Road, Durham DH1 3LE, UK}

\end{center}

\vskip1cm
\begin{abstract}
We study the boundary conditions in 
topologically twisted Chern-Simons matter theories 
with the Lie 3-algebraic structure. 
We find that 
the supersymmetric boundary conditions and 
the gauge invariant boundary conditions 
can be unified as 
complexified gauge invariant boundary conditions 
which lead to supergroup WZW models. 
We propose that the low-energy effective field theories 
on the two-dimensional intersection 
of multiple M2-branes on a holomorphic curve inside K3 
with two non-parallel M5-branes on the K3 
are supergroup WZW models 
from the topologically twisted BLG-model and the ABJM-model. 
\end{abstract}


\newpage
\setcounter{tocdepth}{2}
\tableofcontents

\section{Introduction}
One of the most important clues to understanding M-theory 
is the investigation of the two types of branes, 
namely M2-branes and M5-branes. 
It has been proposed that 
the low-energy dynamics of multiple M2-branes 
probing a flat space is described by 
three-dimensional superconformal Chern-Simons matter theories 
known as the BLG-model
\cite{Bagger:2006sk,Bagger:2007jr,Bagger:2007vi,Gustavsson:2007vu,Gustavsson:2008dy} 
and the ABJM-model 
\cite{Aharony:2008ug}. 
The world-volume theory of the M5-branes 
is believed to be a six-dimensional $\mathcal{N}=(2,0)$ 
superconformal field theory. 
It is much less understood 
due to the lack of a classical Lagrangian description,
although there have been 
many interesting discoveries via its compactification. 
Also the two-dimensional intersection 
of M2-branes with M5-branes 
still remains elusive. 
This brane setup is believed to be one of the 
most promising approaches to the description of the M5-branes 
as it is realized 
when the strongly coupled $(2,0)$ theory is 
away from the conformal fixed point. 

The aim of the present paper is 
to study the two-dimensional intersection of 
multiple M2-branes 
on a supersymmetric two-cycle. In particular, we consider a holomorphic curve inside K3 
with two non-parallel M5-branes on the K3, 
which we will refer to as M5- and M5'-branes. 
We investigate the low-energy effective description 
by starting with the topologically twisted Chern-Simons matter theories 
describing the M2-branes on the holomorphic curve 
and examining the boundary conditions. 
Given the brane configuration of the M2-M5-M5' branes on the K3, 
we determine the boundary conditions for the matter fields as 
supersymmetric boundary conditions 
while we impose those on the gauge fields as 
gauge invariant boundary conditions. 
We find that 
these two different types of boundary conditions 
can be combined into complexified gauge invariant boundary conditions. 
Together with the twisted fermionic fields, 
i.e.\ the spin zero fermions and the spin one fermions, 
we obtain conformal field theories on the Riemann surface 
as the $PSL(2|2)$ WZW action 
from the twisted BLG-model 
and the $GL(N|N)$ WZW action 
from the twisted $U(N)_{k}\times U(N)_{-k}$ ABJM-model.  
We propose that such supergroup WZW models are realized as  
the effective topological theories on the intersection 
of the M2-M5-M5' system on K3, 
which we will call ``topological M-strings''. 

The paper is organized as follows. 
In section \ref{0sec2} 
we present a brane configuration in M-theory on K3 
and establish our setup of the topological M-strings. 
We describe the world-volume theory on the M2-branes 
wrapping a holomorphic curve inside K3 
by performing a partial topological twist 
on the BLG-model \cite{Okazaki:2014sga}. 
In section \ref{0sec3}
we analyze the boundary conditions of 
the topologically twisted BLG theory. 
The matter fields satisfy the supersymmetric boundary conditions 
imposed by the fivebrane 
while the gauge fields obey the gauge invariant boundary conditions 
so that we keep the combined system of the M2-branes \cite{Chu:2009ms}. 
We find the merging of the two boundary conditions 
as complexified gauge invariant boundary conditions. 
In section \ref{0sec4} 
we derive the boundary action 
by taking into account the boundary conditions. 
We argue that the complexified gauge invariant boundary conditions 
lead to the sum of the WZW models for complexified gauge group. 
By putting together the conformally invariant terms 
involving the twisted fermions,  
which are known as symplectic fermions \cite{Kausch:2000fu}, 
i.e.\ fermionic scalar fields and fermionic one-form fields, 
we find the supergroup WZW models. 
We propose that 
the supergroup WZW models are the conformally invariant effective theories 
of the topological M-strings. 
Finally in section \ref{0sec5} 
we close with some discussion.

\section{Topological M-strings}
\label{0sec2}
\begin{figure}
\begin{center}
\includegraphics[width=8cm]{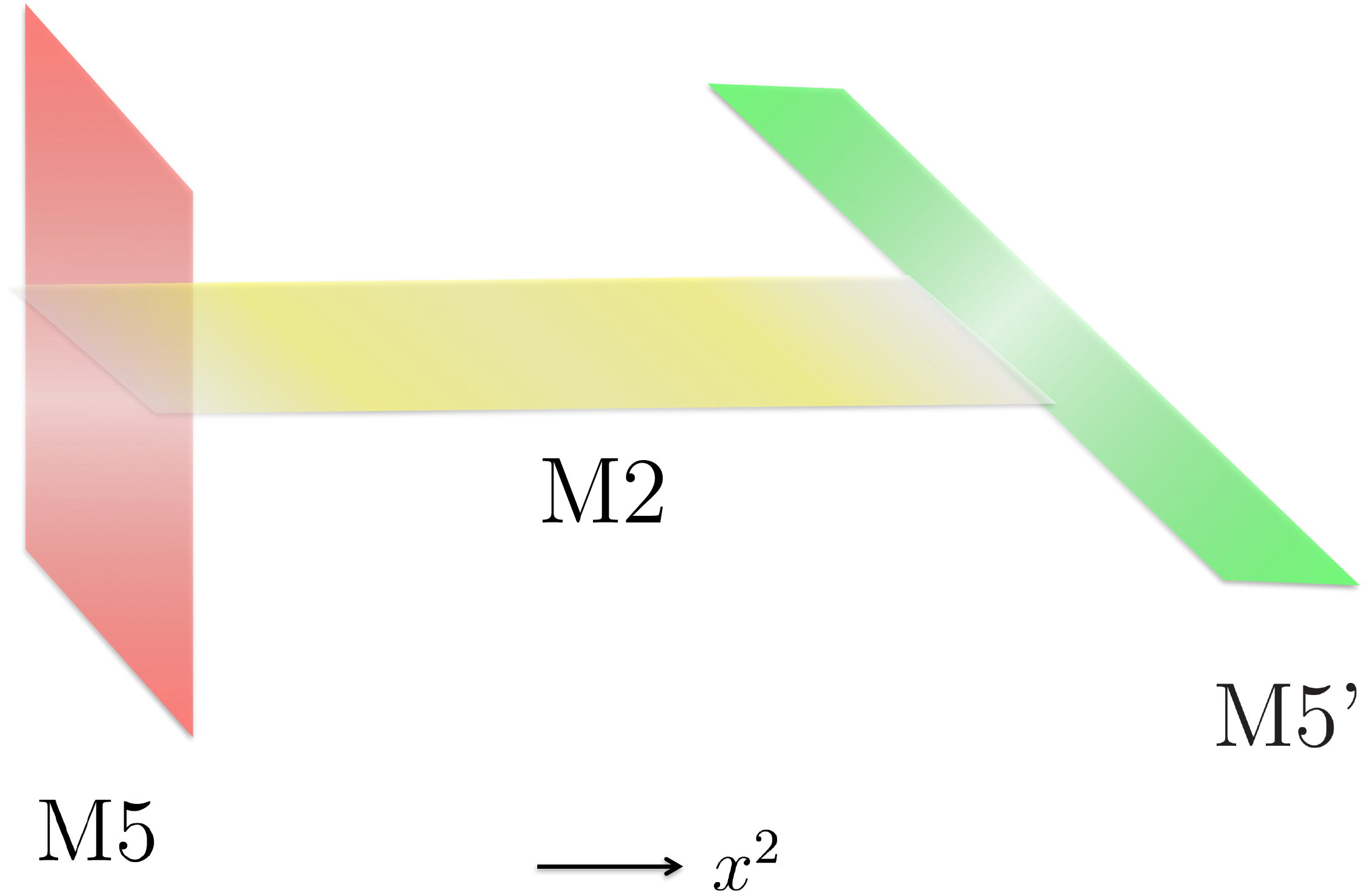}
\caption{M2-branes stretched between 
an M5-brane and an M5'-brane along the $x^{2}$ direction.}
\label{figm2m5b}
\end{center}
\end{figure}

We consider M-theory on the background 
\begin{align}
\label{braneconf1a1}
\mathrm{K3}\times \mathbb{R}^{7}. 
\end{align}
We take the K3 as a cotangent bundle $T^{*}\Sigma_{g}$ 
over a Riemann surface $\Sigma_{g}$ 
where $\Sigma_{g}$ is a holomorphic curve 
of genus $g\neq 1$ in the $x^{0}$, $x^{1}$ directions 
\footnote{
To make the discussion precise, we focus on the case with $g\neq 1$ 
since genus one may require a different treatment for the twisting 
as the surface is flat and 
the supercharges have no charge under the associated flux $F$. 
However, the resulting topologically twisted theory 
would be defined on the surface of genus one.}
.
We take the non-trivial normal bundle $N_{\Sigma}$ 
over the surface in the $x^{9}$, $x^{10}$ directions. 
Let us consider multiple wrapped M2-branes on $\Sigma_{g}\times I$ 
where $I$ is an interval in the $x^{2}$ direction 
with length $L$. 
At one end of the interval 
we put a single fivebrane on $\mathrm{K3}\times \mathbb{R}^{2}_{34}$, 
which we will call an M5-brane,
and at the other end a fivebrane on $\mathrm{K3}\times \mathbb{R}^{2}_{56}$, 
which we will call an M5'-brane. 
The configuration is summarized as
\begin{align}
\label{braneconf1a2}
\begin{array}{cccccccccccc}
&0&1&2&3&4&5&6&7&8&9&10\\
\textrm{M5}
&\circ&\circ& &\circ&\circ& & & & &\circ&\circ\\
\textrm{M5'}
&\circ&\circ& & & &\circ&\circ& & &\circ&\circ\\
\textrm{M2}
&\circ&\circ&\circ& & & & & & & &  \\
\end{array}
\end{align}
and it is depicted in Figure \ref{figm2m5b}. 
%
%

Since the K3 decomposes 
the holonomy group $SO(4)$ of the flat four-manifold into $SU(2)$, 
the spinor representation follows the branching rule  
$\bm{4}$ $\rightarrow$ $\bm{2}\oplus \bm{1}\oplus \bm{1}$ 
and there remains half of the supersymmetry, 
sixteen supercharges in the M-theory background (\ref{braneconf1a1}). 
The presence of the M2-branes, 
the M5-brane and the M5'-brane 
splits the $SO(7)$ Euclidean symmetry group 
into $SO(2)_{34}\times SO(2)_{56}\times SO(2)_{78}$ 
and breaks $1/8$ of the background supersymmetry 
as a consequence of three projections. 
Altogether, 
there are two supercharges preserved 
on the world-volume of the branes.

As the fivebranes are infinite in the directions 
which are not shared by the membranes, 
the fivebranes are much heavier than the membranes. 
Thus the parameters of the fivebranes would be fixed 
and the low-energy effective theory of the branes 
would essentially describe the stretched M2-branes. 
The M2-branes represent minimum energy states 
in a specific topological sector as BPS states. 
We consider the field theory of the membranes 
in which the distance $L$ goes to zero 
and thus it is a two-dimensional sigma model on the intersection. 
The target space of this sigma model 
would be the moduli space of solutions to the BPS constraints 
which encompass the supersymmetric boundary conditions. 
In what follows we will consider 
such effective theories on the M2-branes. 
Recently there have been intriguing approaches 
for the study of M2-branes stretched between M5-branes, 
the so-called M-strings \cite{Haghighat:2013gba}. 
In our brane setup (\ref{braneconf1a2}) 
the M2-branes cannot fluctuate in the flat directions, 
i.e. in $x^{3}$, $x^{4}$, $\cdots$, $x^{8}$, 
so the effective theories on the wrapped M2-branes 
may only capture the topological sector of the M-strings, 
which we call the topological M-strings. 
We will seek the topological sigma model on the intersection  
as the effective theory of the topological M-strings.

The low-energy effective theory 
of curved branes wrapping supersymmetric cycles 
can be obtained by a topological twisting of  
an effective theory of flat branes propagating in a flat space 
\cite{Bershadsky:1995qy}. 
We shall firstly discuss the case of two M2-branes. 
The low-energy effective theory of two coincident membranes propagating 
in a flat space $\mathbb{R}^{8}$ is expected to be realized as the BLG-model 
\cite{Bagger:2006sk,Bagger:2007jr,Bagger:2007vi,Gustavsson:2007vu,Gustavsson:2008dy}. 
The BLG-model is a three-dimensional $\mathcal{N}=8$ superconformal
Chern-Simons matter theory whose action is 
\begin{align}
\label{blg1a1}
S_{\textrm{BLG}}&=S_{\textrm{matter}}+S_{\textrm{TCS}}
\end{align}
where 
\begin{align}
\label{blg1a2}
S_{\textrm{matter}}
&=\int d^{3}x 
\Biggl[
-\frac12 D^{\mu}X^{Ia}D_{\mu}X_{a}^{I}
+\frac{i}{2}\overline{\Psi}^{a}\Gamma^{\mu}D_{\mu}\Psi_{a}\nonumber\\
&
+\frac{i}{4}\overline{\Psi}_{b}\Gamma^{IJ}X^{I}_{c}X^{J}_{d}\Psi_{a}f^{abcd}
-\frac{1}{12}f^{abcd}{f^{efg}}_{d}
X_{a}^{I}X_{b}^{J}X_{c}^{K}
X_{e}^{I}X_{f}^{J}X_{g}^{K}\Biggr]
\end{align}
is the matter action and
\begin{align}
\label{blg1a3}
S_{\textrm{TCS}}
&=\frac12\epsilon^{\mu\nu\lambda}
\left(
f^{abcd}A_{\mu ab}\partial_{\nu}A_{\lambda cd}
+\frac23 {f^{cda}}_{g}f^{efgb}
A_{\mu ab}A_{\nu cd}A_{\lambda ef}
\right)
\end{align}
is the twisted Chern-Simons action 
in terms of the structure constants ${f^{abc}}_{d}$ 
of the Lie 3-algebra. 
Only the $\mathcal{A}_{4}$ algebra 
with ${f^{abc}}_{d}=\frac{2\pi}{k}\epsilon^{abcd}$, 
$k\in \mathbb{Z}$, $a,b,\cdots,=1,\cdots,4$
admits a finite dimensional non-trivial representation of 
the Lie 3-algebra with a positive definite metric. 
The field content is 
eight real scalar fields $X^{I}_{a}$, $I=1,\cdots,8$ 
describing the position of the M2-branes 
in the flat eight-dimensional space, 
fermionic fields $\Psi_{a}$ defined as the $SO(1,10)$ Majorana spinor 
obeying the projection 
$\Gamma_{012}\Psi=-\Psi$ and gauge fields $A_{\mu ab}$, $\mu=0,1,2$ 
where the gauge indices $a,b,\cdots$ run from $1$ to $4$. 
The theory has a three-dimensional Lorentz group $SO(1,2)$ 
as the rotational symmetry group on the world-volume of the membranes 
and the R-symmetry group $SO(8)_{R}$ 
as the isometry of the transverse space of the membranes. 
The fields $X^{I}_{a}$, $\Psi_{a}$ and $A_{\mu ab}$ 
transform under the $SO(1,2)\times SO(8)_{R}$ 
as $(\bm{1},\bm{8}_{v})$, $(\bm{2},\bm{8}_{c})$ 
and $(\bm{3},\bm{1})$ respectively. 
The action (\ref{blg1a1}) is invariant under the supersymmetry transformations
\begin{align}
\label{blg1b1}
\delta X_{a}^{I}
&=i\overline{\epsilon}\Gamma^{I}\Psi_{a}\\
\label{blg1b2}
\delta\Psi_{a}
&=D_{\mu}X_{a}^{I}\Gamma^{\mu}\Gamma^{I}\epsilon
-\frac16 X_{b}^{I}X_{c}^{J}X_{d}^{K}{f^{bcd}}_{a}\Gamma^{IJK}\epsilon\\
\label{blg1b3}
{{\delta\tilde{A}_{\mu}}^{b}}_{a}
&=i\overline{\epsilon}\Gamma_{\mu}\Gamma^{I}X_{c}^{I}\Psi_{d}{f^{cdb}}_{a}
\end{align}
where we have defined 
${\tilde{A}_{\mu}}{\ ^{b}}_{a}:={f^{cdb}}_{a}A_{\mu cd}$. 
The supersymmetry parameter $\epsilon$ 
is the $SO(1,10)$ Majorana spinor 
satisfying the projection 
$\Gamma_{012}\epsilon=\epsilon$ 
and transforms as $(\bm{2},\bm{8}_{s})$.

In order to study the two wrapped M2-branes 
on a holomorphic Riemann surface $\Sigma_{g}$, 
we consider a partial topological twisting of the BLG-model. 
Such a topological twisting replaces the Euclidean symmetry group
$SO(2)_{E}$ of the two-dimensional space 
by a different subgroup $SO(2)_{E}'$ of the 
$SO(2)_{E}\times SO(8)_{R}$ 
so that there exist scalar supercharges 
as discussed in \cite{Okazaki:2014sga}. 
There are plenty of twists 
by taking a homomorphism $\mathfrak{h}$: $SO(2)\rightarrow SO(8)_{R}$. 
The partial topological twisting 
for the M2-branes wrapped on a holomorphic curve 
inside K3 can be uniquely determined 
by decomposing the $SO(8)_{R}$ $\rightarrow$ $SO(2)_{R}\times SO(6)_{R}$ 
and defining $SO(2)'_{E}$ 
$=$ 
$\mathrm{diag}(SO(2)_{E}\times SO(2)_{R})$. 
After the twist the bosonic matter fields transform 
under $SO(2)_{E}'\times SO(6)_{R}$ as
\begin{align}
\label{twist1a1}
\bm{6}_{0}\oplus \bm{1}_{2}\oplus\bm{1}_{-2}
\end{align}
and one obtains 
the resultant bosonic scalar fields $\bm{6}_{0}$ 
which we denote by $\phi^{I}$, where now $I=1,\cdots,6$ 
and a bosonic one-form $\bm{1}_{2}$ and $\bm{1}_{-2}$ 
which we denote by $\Phi_{z}$ and $\Phi_{\overline{z}}$. 
The representation of the fermionic fields is
\begin{align}
\label{twista2}
\bm{4}_{2}\oplus \overline{\bm{4}}_{0}
\oplus \bm{4}_{0}\oplus \overline{\bm{4}}_{-2}
\end{align}
where $\bm{4}_{0}$, $\overline{\bm{4}}_{0}$ are 
the fermionic scalar fields which we will denote by $\psi$,
$\tilde{\lambda}$ 
while $\bm{4}_{2}$, $\overline{\bm{4}}_{-2}$ 
are the fermionic one-form fields which we will denote by $\Psi_{z}$,
$\tilde{\Psi}_{\overline{z}}$.  
The supersymmetry parameter $\epsilon$ transforms as
\begin{align}
\label{twista3}
\bm{4}_{0}\oplus\overline{\bm{4}}_{2}
\oplus\bm{4}_{-2}\oplus \overline{\bm{4}}_{0}
\end{align}
under $SO(2)_{E}'\times SO(6)$ 
and thus one can find eight scalar supercharges 
associated with the supersymmetric parameters 
$\xi$, $\tilde{\xi}$ 
in the representation $\bm{4}_{0}\oplus \overline{\bm{4}}_{0}$. 
Note that topological twisting modifies the original theory 
so that the new symmetry group $SO(2)_{E}'$ is defined by 
the diagonal subgroup of $SO(2)_{E}$ and $SO(2)_{R}=SO(2)_{910}$ 
as diag$(SO(2)_{E}\times SO(2)_{910})$. 
In other words, 
the new generator of $SO(2)'_{E}$ has been created by 
the generator of $SO(2)_{E}$ and that of $SO(2)_{910}$. 
So we could only say that the resulting twisted theory contains the
modified $SO(2)_{E}'$ symmetry group.

From the M-theory point of view, 
a compactification on K3 retains  
seven flat directions and six of them 
are transverse to the membranes' world-volume 
$\Sigma_{g}\times I$. 
Thus we see that 
the above topological twisting exactly realizes 
the required bosonic field content in the effective theory 
of the M2-branes wrapped on $\Sigma_{g}$ inside K3.
The six bosonic scalars $\phi^{I}$ 
describe the displacement of the M2-branes in the six flat directions 
while the bosonic one-form field $\Phi_{\alpha}$ on the Riemann surface 
describes the motion of the M2-branes inside K3, 
i.e. the non-trivial normal bundle $N_{\Sigma}$ over $\Sigma_{g}$. 
The existence of eight covariantly constant spinors in (\ref{twista3}) 
reflects the fact that K3 breaks half of the supersymmetry.

\section{Boundary Conditions}
\label{0sec3}

\subsection{Supersymmetric boundary conditions}
\label{secSUSYbc}
To extend the study of the compact M2-branes 
wrapped around a holomorphic curve inside K3 
to the M2-M5-M5' system (\ref{braneconf1a2}),  
we will analyze the boundary conditions 
which should be imposed by the M5-brane and the M5'-brane 
at the ends of the M2-branes in the $x^{2}$ direction. 
Let us start our investigation 
by considering maximally supersymmetric boundary conditions, 
i.e. half-BPS boundary conditions of the BLG-model 
which include the case where the M2-branes end on an M5-brane
\footnote{See \cite{Berman:2009xd} for more general discussions.}. 

The supersymmetry is preserved on the boundary 
when the normal component of the supercurrent vanishes on the boundary 
\cite{Gaiotto:2008sa,Berman:2009xd,Okazaki:2013kaa,Chung:2016pgt}. 
The supersymmetric transformations (\ref{blg1b1})-(\ref{blg1b3}) 
lead to a supercurrent 
\begin{align}
\label{scurrent1}
J^{\mu}&=
-D_{\nu}X^{Ia}\Gamma^{\nu}\Gamma^{I}\Gamma^{\mu}\Psi_{a}
-\frac16 X_{a}^{I}X_{b}^{J}X_{c}^{K}f^{abcd}\Gamma^{IJK}\Gamma^{\mu}\Psi_{d} .
\end{align}
Let the M2-branes with world-volume $(x^{0},x^{1},x^{2})$ 
end on a single M5-brane with world-volume 
$(x^{0},x^{1},x^{3},x^{4},x^{9},x^{10})$ at, say, $x^{2}=0$. 
According to the existence of the M5-brane 
the $SO(8)_{R}$ splits into 
$SO(4)_{34910}\times SO(4)_{5678}$. 
Correspondingly we will decompose the eight scalar fields 
into two parts; 
$X^{i}$ $=$ $\left\{X^{3},X^{4},X^{9},X^{10}\right\}$, 
$Y^{\hat{i}}$ $=$ $\left\{ X^{5},X^{6},X^{7},X^{8}\right\}$. 
Then the supersymmetric boundary conditions can be written as 
\begin{align}
\label{susybc1a1}
0&=\overline{\epsilon}J^{2}\Big|_{\textrm{bdy}}\nonumber\\
&=D_{\alpha}X^{i}
\left(
\overline{\epsilon}\Gamma^{i\alpha 2}\Psi
\right)
+
D_{\alpha}Y^{\hat{i}}
\left(
\overline{\epsilon}\Gamma^{\hat{i}\alpha 2}\Psi
\right)\nonumber\\
&+D_{2}X^{i}
\left(
\overline{\epsilon}\Gamma^{i}\Psi
\right)
+D_{2}Y^{\hat{i}}
\left(
\overline{\epsilon}\Gamma^{\hat{i}}\Psi
\right)\nonumber\\
&-\frac16 \left[
X^{i},X^{j},X^{k}
\right]
\left(
\overline{\epsilon}\Gamma^{ijk}\Gamma^{2}\Psi
\right)
-\frac16 \left[
Y^{\hat{i}},Y^{\hat{j}},Y^{\hat{k}}
\right]
\left(
\overline{\epsilon}\Gamma^{\hat{i}\hat{j}\hat{k}}\Gamma^{2}\Psi
\right)\nonumber\\
&-\frac12 \left[
X^{i},X^{j},Y^{\hat{k}}
\right]
\left(
\overline{\epsilon}\Gamma^{ij\hat{k}}\Gamma^{2}\Psi
\right)
-\frac12 \left[
X^{i},Y^{\hat{j}},Y^{\hat{k}}
\right]
\left(
\overline{\epsilon}\Gamma^{i\hat{j}\hat{k}}\Gamma^{2}\Psi
\right)
\Bigg|_{\textrm{bdy}}.
\end{align}

To find the solutions to the half-BPS boundary conditions 
which correspond to the M2-M5 system, 
we should take into account 
several constraints from the brane configuration. 
We observe that 
all the parameters of the M5-brane are fixed 
so that the scalar fields $Y^{\hat{i}}$ should 
obey the Dirichlet boundary conditions $D_{\alpha}Y^{\hat{i}}=0$. 
Since we do not expect the M5-brane to impose both Neumann and
Dirichlet boundary conditions on the scalar fields $Y^{\hat{i}}$,
$D_2Y^{\hat{i}}=0$ should not also be constrained at the boundary. Therefore, in order
to satisfy the boundary condition (\ref{susybc1a1})
we also need to choose appropriate boundary conditions 
for the fermionic fields. 

Noting that 
the unbroken supersymmetry parameter $\epsilon$ 
must satisfy the projections
$\Gamma_{0134910}\epsilon=\epsilon$ due to the M5-brane and
$\Gamma_{012}\epsilon=\epsilon$ due to the M2-branes,
one finds
\begin{align}
\label{pro1a1}
\Gamma_{01}\epsilon
=\Gamma_{2}\epsilon
=\Gamma_{34910}\epsilon
=\Gamma_{5678}\epsilon.
\end{align}
%
%
There remain eight supercharges on the two-dimensional boundary. 
In two dimensions supersymmetry has a definite chirality. 
Equation (\ref{pro1a1}) shows that 
the chiralities of the supersymmetry parameter 
under the $SO(1,1)_{01}$, 
the 
$SO(4)_{34910}$ 
and the 
$SO(4)_{5678}$ 
are the same. 
%
%
%
%
It is easily checked that $\Gamma_{01}$ is Hermitian, traceless and squares
to the identity on the eight-dimensional subspace of spinors satisfying
equation (\ref{pro1a1}).
Thus $\mathcal{N} = (4,4)$ supersymmetry is preserved on the two-dimensional boundary 
of the M2-branes ending on the M5-brane. 

As in the projection conditions (\ref{pro1a1}), 
we can employ the boundary condition ansatz for the fermions 
$\Gamma_{0134910}\Psi=\Psi$ 
so that 
the space-time symmetry of the brane configuration is maintained. 
Combining the boundary conditions (\ref{susybc1a1}) 
with the fermionic boundary conditions satisfying 
the restrictions from the M2-M5 configuration 
we get the half-BPS boundary conditions at the M5-brane 
\begin{align}
\label{susybc1a2}
D_{2}X^{i}+\frac16 \epsilon^{ijkl}[X^{j},X^{k},X^{l}]
\Bigl|_{\textrm{bdy}}&=0,\\
\label{susybc1a3}
D_{\alpha}Y^{\hat{i}}\Bigl|_{\textrm{bdy}}&=0
\end{align}
where $\epsilon^{ijkl}$ is an antisymmetric tensor with
$\epsilon^{34910}=1$
\footnote{There is also a third condition imposed,
$[X^i, Y^{\hat{j}}, Y^{\hat{k}}] = 0$ but we could simply set
$Y^{\hat{i}} = 0$ on the M5-brane, noting the
Dirichlet boundary conditions (\ref{susybc1a3}).}
. 
The first boundary condition (\ref{susybc1a2}) 
is the Basu-Harvey equation \cite{Basu:2004ed} 
which would describe the displacement of the M2-branes 
in the four-dimensional space inside the M5-brane. 
The second boundary condition (\ref{susybc1a3}) 
fixes the boundary values of the position of the M2-branes 
in the remaining four-dimensional space which is normal to the M5-brane.

One can easily obtain  
the boundary conditions from an additional fivebrane. 
Consider the fivebrane 
with world-volume $(x^{0},x^{1},x^{5},x^{6},x^{7},x^{8})$, 
which we will denote by $\widetilde{\textrm{M5}}$-brane. 
By exchanging a role of $X^{i}$ and $Y^{\hat{i}}$, 
we obtain the boundary conditions 
\begin{align}
\label{susybc1a4}
D_{2}Y^{\hat{i}}+\frac16 \epsilon^{\hat{i}\hat{j}\hat{k}\hat{l}}
[Y^{\hat{j}},Y^{\hat{k}},Y^{\hat{l}}]
\Bigl|_{\textrm{bdy}}&=0,\\
\label{susybc1a5}
D_{\alpha}X^{i}\Bigl|_{\textrm{bdy}}&=0.
\end{align}
Adding the boundary conditions (\ref{susybc1a4}) and (\ref{susybc1a5}) 
from the 
$\widetilde{\textrm{M5}}$-brane 
to the boundary conditions (\ref{susybc1a2}) and (\ref{susybc1a3})
from the M5-brane 
does not break further supersymmetry. 
%
%
So the effective theories of the intersecting 
M2-M5-$\widetilde{\textrm{M5}}$ 
system in a flat space 
would be 
two-dimensional $\mathcal{N}=(4,4)$ superconformal field theories. 
Although the knowledge of such 
field theories is still limited, 
the 
M2-M5-$\widetilde{\textrm{M5}}$ 
solutions whose near-horizon geometries take the form 
AdS$_{3}\times$ $S^{3}\times$ $S^{3}$ 
have been constructed in the gravity dual perspective 
\cite{Boonstra:1998yu,deBoer:1999rh,Bachas:2013vza}.  
%
%
%
%
%
%

%
%
Let us instead consider 
the flat M5'-brane located along $(x^{0},x^{1},x^{5},x^{6},x^{9},x^{10})$ 
having four common directions with the M5-brane. 
The isometry of the transverse space of the M2-branes 
reduces to $SO(2)_{34}\times SO(2)_{56}\times SO(2)_{78}\times
SO(2)_{910}$. 
We thus decompose the eight scalar fields as 
$X^{i}=(X^{9},X^{10})$, 
$Y^{\hat{i}}=(X^{7},X^{8})$, 
$Z^{i}=(X^{3},X^{4})$ 
and 
$\hat{Z}^{\hat{i}}=(X^{5},X^{6})$. 
The preserved supersymmetry parameters $\epsilon$ 
should satisfy 
$\Gamma_{012}\epsilon=\epsilon$, 
$\Gamma_{0134910}\epsilon=\epsilon$ and 
$\Gamma_{0156910}\epsilon=\epsilon$, 
from which one can read 
their chiralities under the 
$SO(1,1)_{01}\times 
SO(2)_{34}\times SO(2)_{56}\times SO(2)_{78}\times
SO(2)_{910}$ 
as 
$(+,+,+,-,-)$, 
$(+,-,-,+,+)$, 
$(-,+,+,+,+)$ and 
$(-,-,-,-,-)$. 
Thus $\mathcal{N}=(2,2)$ supersymmetry 
is preserved on the two-dimensional intersection 
of the M2-M5-M5' system. 

Adopting the fermionic boundary conditions 
$\Gamma_{0134910}\Psi = \Psi$ on the M5-brane and
$\Gamma_{0156910}\Psi=\Psi$ on the M5'-brane, in the limit where the
M5-M5' separation is small,
we find the common set of boundary conditions on the bosonic fields 
\begin{align}
\label{m2m5m5bc01a}
&D_{2}X^{i}
+\frac12 \epsilon^{ijkl}
[X^{j},Z^{k},Z^{l}]
+\frac12 \epsilon^{ij\hat{k}\hat{l}}
[X^{j},\hat{Z}^{\hat{k}},\hat{Z}^{l}]=0,\\
\label{m2m5m5bc01b}
&D_{\alpha}Y^{\hat{i}}=0,\\
&D_{\alpha}Z^{i}=0,\ \ \ \ \ 
D_{2}Z^{i}+\frac12 \epsilon^{ijkl}[Z^{j},X^{k},X^{l}]=0,\\
&D_{\alpha}\hat{Z}^{\hat{i}}=0,\ \ \ \ \ 
D_{2}\hat{Z}^{\hat{i}}
+\frac12\epsilon^{\hat{i}\hat{j}kl}[\hat{Z}^{\hat{j}},X^{k},X^{l}]=0.
\end{align}
The first equation (\ref{m2m5m5bc01a}) is the Basu-Harvey like equation 
for $X^{i}$ with 
two of the elements in the three bracket replaced 
by $Z^{i}$ or $\hat{Z}^{\hat{i}}$. 
The second equation (\ref{m2m5m5bc01b}) 
is the Dirichlet boundary condition on $Y^{\hat{i}}$. 
The last two equations are curious 
since the $Z^{i}$ and $\hat{Z}^{\hat{i}}$ 
are required to be fixed at the boundary by one of the fivebranes 
while they should also keep the Lie 3-algebraic structure 
due to the non-vanishing three-bracket. 
Although the direct analysis of the $\mathcal{N}=(2,2)$ 
superconformal field theories is still difficult, 
their supersymmetric ground states, 
the chiral rings, the BPS spectra 
and the sphere partition functions 
have been investigated by taking the mass deformation 
in \cite{Hori:2013ewa}. 

Now we will proceed to the boundary conditions 
in the topologically twisted BLG theory 
describing the curved M2-branes. 
Let us decompose the $SO(1,10)$ gamma matrices as 
\begin{equation}
\label{mtx1}
\begin{cases}
 \Gamma^{\mu}=\gamma^{\mu}\otimes
  \hat{\Gamma}^{7}
\otimes \sigma_{2}
&\mu=0,1,2 \cr
 \Gamma^{I+2}=\mathbb{I}_{2}\otimes
 \hat{\Gamma}^{I} \otimes 
\sigma_{2}&I=1,\cdots, 6\cr
 \Gamma^{i+8}=\mathbb{I}_{2}\otimes
 \mathbb{I}_{8}\otimes
 \gamma^{i}&i=1,2 \cr
\end{cases}
\end{equation}
where $\gamma^{\mu}$, $\mu=0,1,2$ are $2\times 2$ matrices; 
$\gamma^{0}=\sigma_{1}$, 
$\gamma^{1}=\sigma_{3}$ and 
$\gamma^{2}=i\sigma_{2}$, while
$\hat{\Gamma}^{I}$ are the $SO(6)$ gamma matrices satisfying the relations 
\begin{align}
\label{gamma0a2}
\{\hat{\Gamma}^{I},\hat{\Gamma}^{J}\}=2\delta^{IJ},\ \ \ 
(\hat{\Gamma}^{I})^{\dag}=\Gamma^{I},
\end{align}
\begin{align}
\label{gamma7}
\hat{\Gamma}^{7}
=-i\hat{\Gamma}^{12\cdots 6}=\left(
\begin{array}{cc}
\mathbb{I}_{4}&0\\
0&-\mathbb{I}_{4}\\
\end{array}
\right).
\end{align}
The $SO(1,10)$ charge conjugation matrix $\mathcal{C}$ is decomposed as
\begin{align}
\label{ccmtx}
\mathcal{C}=\epsilon \otimes \hat{C} \otimes \epsilon
\end{align}
where the $SO(2)$ charge conjugation matrix $\epsilon$ 
and the $SO(6)$ charge conjugation matrix $\hat{C}$ obey the relations
\begin{align}
\label{ccmtx0}
\epsilon^{T}=-\epsilon,\ \ \
\epsilon\gamma^{\mu}\epsilon^{-1}=-(\gamma^{\mu})^{T}, 
\end{align}
\begin{align}
\label{ccmtx1}
\hat{C}^{T}=-\hat{C},\ \ \ \ \ \ \hat{C}\hat{\Gamma}^{I}\hat{C}^{-1}=(\hat{\Gamma}^{I})^{T},\ \ \ \ \ \ 
\hat{C}\hat{\Gamma}^{7}\hat{C}^{-1}=-(\hat{\Gamma}^{7})^{T}.
\end{align}
%
We will write the twisted bosonic fields as
\begin{align}
\label{new1}
\phi^{I}&:=X^{I+2},\\
\label{new2}
\Phi_{z}&:=\frac{1}{\sqrt{2}}(X^{9}-iX^{10}), 
&\Phi_{\overline{z}}&:=\frac{1}{\sqrt{2}}(X^{9}+iX^{10}),\\
\label{new3}
A_{z}&:=\frac{1}{\sqrt{2}}(A_{1}-iA_{2}), 
&A_{\overline{z}}&:=\frac{1}{\sqrt{2}}(A_{1}+iA_{2}).
\end{align}
To treat the twisted fermionic fields  
we expand them as 
\begin{align}
\label{new4}
\Psi_{A}^{\alpha\beta}
=\frac{i}{\sqrt{2}}
\psi_{A}(\gamma_{+}\epsilon^{-1})^{\alpha\beta}
+i\tilde{\Psi}_{\overline{z}A}
(\gamma^{\overline{z}}\epsilon^{-1})^{\alpha\beta}
-\frac{i}{\sqrt{2}}\tilde{\lambda}_{A}
(\gamma_{-}\epsilon^{-1})^{\alpha\beta}
-i\Psi_{zA}(\gamma^{z}\epsilon^{-1})^{\alpha\beta}
\end{align}
where we have introduced the $2\times 2$ 
matrices $\gamma_{\pm}, \gamma^{z}$ and $\gamma^{\overline{z}}$ defined by 
\begin{align}
\label{gammapm}
&\gamma_{+}:=\frac{1}{\sqrt{2}}(\mathbb{I}_{2}+\sigma_{2}),\ \ \ \ \ 
\gamma_{-}:=\frac{1}{\sqrt{2}}(\mathbb{I}_{2}-\sigma_{2}),\\
\label{gammaz1}
&\gamma^{z}
:=\frac{1}{\sqrt{2}}
(\gamma^{1}+i\gamma^{2})=\frac{1}{\sqrt{2}}\left(
\begin{array}{cc}
i&1\\
1&-i\\
\end{array}
\right), \\
\label{gammaz2}
&\gamma^{\overline{z}}
:=\frac{1}{\sqrt{2}}(\gamma^{1}-i\gamma^{2})
=\frac{1}{\sqrt{2}}\left(
\begin{array}{cc}
-i&1\\
1&i\\
\end{array}
\right)
\end{align}
and the indices $\alpha,A$ and $\beta$ label 
the $SO(2)_{E}$ spinor,
the $SO(6)_{R}$ spinor 
and the 
{$SO(2)_{R}$}  
spinor respectively. 
The supersymmetry parameter can be expanded in a similar fashion
\begin{align}
\label{new5}
\epsilon_{A}^{\alpha\beta}
=\frac{i}{\sqrt{2}}\tilde{\xi}_{A}
(\gamma_{+}\epsilon^{-1})^{\alpha\beta}
+i\epsilon_{\overline{z}A}
(\gamma^{\overline{z}}\epsilon^{-1})^{\alpha\beta}
-\frac{i}{\sqrt{2}}\xi_{A}
(\gamma_{-}\epsilon^{-1})^{\alpha\beta}
-i\tilde{\epsilon}_{zA}
(\gamma^{z}\epsilon^{-1})^{\alpha\beta}.
\end{align}
Note that only 
$\xi$ and $\tilde{\xi}$ play the role of 
supersymmetry parameters on $\Sigma_{g}$ 
as they behave as covariantly constant spinors.

\begin{figure}
\begin{center}
\includegraphics[width=3cm]{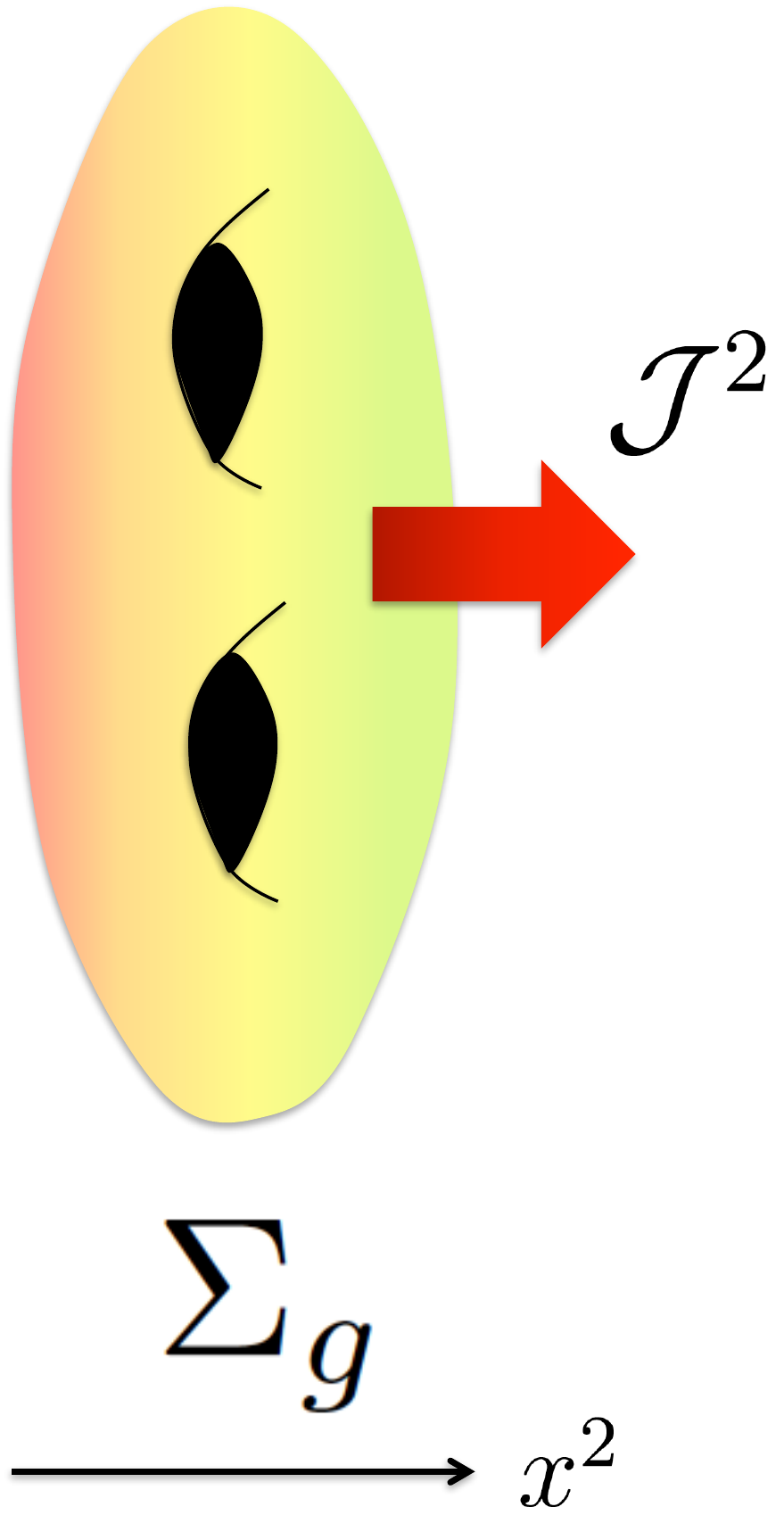}
\caption{The BRST current $\mathcal{J}^{2}$ in the effective theory 
of the M2-branes wrapping $\Sigma_{g}$. 
The current is associated with the BRST charges defined on
 $\Sigma_{g}$. 
We require that it vanishes at the boundary.}
\label{figm2m5c}
\end{center}
\end{figure}

Using the expressions defined above, 
we find the supersymmetric boundary conditions 
in the twisted BLG theory
\begin{align}
\label{susybc2a1}
0=&
\overline{\xi}\mathcal{J}^{2}
-\overline{\tilde{\xi}}\tilde{\mathcal{J}}^{2}
\Bigl|_{\textrm{bdy}}
\nonumber\\
=&-\overline{\xi}
\left[
D_{2}\phi^{I}\hat{\Gamma}^{I}
-\frac16 [\phi^{I},\phi^{J},\phi^{K}]\hat{\Gamma}^{IJK}
-[\Phi_{z},\Phi_{\overline{z}},\phi^{I}]\hat{\Gamma}^{I}
\right]\psi
\nonumber\\
&-\overline{\xi}
\left[
iD_{\overline{z}}\phi^{I}\hat{\Gamma}^{I}
\right]\Psi_{z}
-\overline{\xi}
\left[
2iD_{\overline{z}}\Phi_{z}
\right]\tilde{\lambda}
-\overline{\xi}
\left[
D_{2}\Phi_{z}+\frac12 [\phi^{I},\phi^{J},\Phi_{z}]\hat{\Gamma}^{IJ}
\right]\tilde{\Psi}_{\overline{z}}\nonumber\\
&+\overline{\tilde{\xi}}
\left[
D_{2}\phi^{I}\hat{\Gamma}^{I}
-\frac16 [\phi^{I},\phi^{J},\phi^{K}]\hat{\Gamma}^{IJK}
-[\Phi_{z},\Phi_{\overline{z}},\phi^{I}]\hat{\Gamma}^{I}
\right]\tilde{\lambda}\nonumber\\
&+\overline{\tilde{\xi}}
\left[
iD_{z}\phi^{I}\hat{\Gamma}^{I}
\right]\tilde{\Psi}_{\overline{z}}
+\overline{\tilde{\xi}}
\left[
2iD_{z}\Phi_{\overline{z}}
\right]\psi
+\overline{\tilde{\xi}}
\left[
-D_{2}\Phi_{\overline{z}}
-\frac12 [\phi^{I},\phi^{J},\Phi_{\overline{z}}]\hat{\Gamma}^{IJ}
\right]\Psi_{z}\Biggl|_{\textrm{bdy}}
\end{align}
where $\mathcal{J}^{2}$ 
($\tilde{\mathcal{J}}^{2}$) 
is the BRST current associated with 
the supersymmetry parameter $\xi$ 
$(\tilde{\xi})$
transforming as a scalar on
$\Sigma_{g}$ (see Figure \ref{figm2m5c}). 

Although there would be 
various solutions to the equation (\ref{susybc2a1}), 
we are interested in the solutions which correspond to 
the M2-M5-M5' system (\ref{braneconf1a2}). 
We can apply the general lesson 
we have learned in the flat case to find them. 
At the boundary of the M5-brane 
the bosonic one-form field $\Phi_{\alpha}$ 
should obey a particular boundary condition.
However, the bosonic scalar fields $\phi^{I}$ 
should have two different types of boundary conditions 
due to the tangent and normal directions 
of the attached M5-brane.
These are expected to be the Basu-Harvey like boundary condition 
describing a non-trivial geometry inside the M5-brane  
and the Dirichlet boundary condition respectively. 
Let $\phi^{\hat{i}}$, $\hat{i}=1,2$ be the scalar fields 
$\phi^{1}$ and $\phi^{2}$
which represent the position of the M2-branes 
within the M5-brane
and let $\rho^{\hat{p}}$, $\hat{p}=1,\cdots,4$ be 
$\phi^{3}$, $\cdots$, $\phi^{6}$
which correspond to the transverse directions of the M5-brane. 
To obtain the Dirichlet condition on $\rho^{\hat{p}}$ 
we must require the fermionic boundary conditions 
\begin{align}
\label{susybc2a2}
\overline{\xi}\hat{\Gamma}^{\hat{p}+4}\psi
\Bigl|_{\textrm{bdy}}&=0, 
&\overline{\tilde{\xi}}\hat{\Gamma}^{\hat{p}+4}\tilde{\lambda}
\Bigl|_{\textrm{bdy}}&=0.
\end{align}
On the other hand, 
the Basu-Harvey type condition on $\phi^{\hat{i}}$ 
can be acquired by choosing the fermionic boundary conditions 
\begin{align}
\label{susybc2a3}
\overline{\xi}\hat{\Gamma}^{\hat{i}+2}\Psi_{z}
\Bigl|_{\textrm{bdy}}&=0, &
\overline{\tilde{\xi}}\hat{\Gamma}^{\hat{i}+2}\tilde{\Psi}_{\overline{z}}
\Bigl|_{\textrm{bdy}}&=0.
\end{align}
The set of equations (\ref{susybc2a2}) and (\ref{susybc2a3}) 
states that 
the fermion bilinear forms cannot play the role of 
generators of translations in the corresponding directions.

Employing the fermionic boundary conditions 
(\ref{susybc2a2}) and (\ref{susybc2a3}), 
we can read off from the generic supersymmetric condition 
(\ref{susybc2a1}) 
the boundary conditions 
at the intersection of the M2-branes and the M5-brane 
in the brane configuration (\ref{braneconf1a2})
\begin{align}
\label{susybc2a4}
D_{2}\phi^{\hat{i}}-[\Phi_{z},\Phi_{\overline{z}},\phi^{\hat{i}}]
\Bigl|_{\textrm{bdy}}&=0,\\
\label{susybc2a4a}
D_{\overline{z}}\rho^{\hat{p}}
\Bigl|_{\textrm{bdy}}&=0, \ \ \ 
D_{z}\rho^{\hat{p}}
\Bigl|_{\textrm{bdy}}=0,\\
\label{susybc2a4b}
D_{\overline{z}}\Phi_{z}
\Bigl|_{\textrm{bdy}}&=0, \ \ \ 
D_{z}\Phi_{\overline{z}}
\Bigl|_{\textrm{bdy}}=0,\\
\label{susybc2a4c}
\overline{\xi}
\left(
D_{2}\Phi_{z}+\frac12 [\rho^{\hat{p}},\rho^{\hat{q}},\Phi_{z}]
\hat{\Gamma}^{\hat{p}\hat{q}}
\right)\tilde{\Psi}_{\overline{z}}
\Bigl|_{\textrm{bdy}}&=0, 
\ \ \ 
\overline{\tilde{\xi}}
\left(
D_{2}\Phi_{\overline{z}}+\frac12 [\rho^{\hat{p}},\rho^{\hat{q}},
\Phi_{\overline{z}}]\hat{\Gamma}^{\hat{p}\hat{q}}
\right)\Psi_{z}
\Bigl|_{\textrm{bdy}}=0.
\end{align}
%
%
The equation (\ref{susybc2a4}) is the 
Basu-Harvey like equation on the scalars $\phi^{\hat{i}}$ 
and the set of equations (\ref{susybc2a4a}) 
is the Dirichlet boundary condition on the scalars $\rho^{\hat{p}}$. 
Note that the set of equations (\ref{susybc2a4b}) 
is not the Dirichlet boundary condition, 
but the holomorphic and anti-holomorphic boundary conditions 
on the one-form fields $\Phi_{z}$ and $\Phi_{\overline{z}}$ 
which are complex-valued functions on the Riemann surface $\Sigma_{g}$. 
Consequently $\Phi_{z}dz$ is a holomorphic differential one-form 
while $\Phi_{\overline{z}}d\overline{z}$ is an anti-holomorphic
differential one-form on $\Sigma_{g}$. 
The field $\Phi_{z}$ satisfying equation (\ref{susybc2a4b}) 
describes a choice of the holomorphic curve $\Sigma_{g}$ in K3.

Likewise we can find the boundary conditions 
at the M5'-brane by 
exchanging the pair of directions $(x^{3},x^{4})$ 
with $(x^{5},x^{6})$. 
Putting the M5'-brane in the M2-M5 configuration 
breaks down the space-time symmetry group 
$SO(4)_{3456}$ to $SO(2)_{34}\times SO(2)_{56}$ 
while it maintains the preserved supersymmetry, as we will see momentarily. 
Let $\varphi^{\hat{i}}$, $i=1,2$ be the bosonic scalars  
which correspond to the position of the M2-brane in the $(x^{3},x^{4})$, 
$\varsigma^{\hat{l}}$, $\hat{l}=1,2$ be those in the $(x^{5},x^{6})$ 
and $\varrho^{\hat{p}}$, $\hat{p}=1,2$ be those in the $(x^{7},x^{8})$. 
The first two, $\varphi^{\hat{i}}$ and $\varsigma^{\hat{l}}$ 
should obey the Basu-Harvey type condition 
as they probe in one of the fivebranes 
while the third $\varrho^{\hat{p}}$ must be subject to the Dirichlet
condition. 
Now consider the limit in which the distance $L$ goes to zero, 
the intersection of the M2-M5-M5' branes (\ref{braneconf1a2}). 
The boundary conditions can be determined 
by combining the two types of conditions required from 
M5-brane and M5'-brane as
\begin{align}
\label{susybc2a5}
D_{2}\varphi^{\hat{i}}-[\Phi_{z},\Phi_{\overline{z}},\varphi^{\hat{i}}]
\Bigl|_{\textrm{bdy}}&=0, 
&D_{z}\varphi^{\hat{i}}
\Bigl|_{\textrm{bdy}}&=0, 
&D_{\overline{z}}\varphi^{\hat{i}}
\Bigl|_{\textrm{bdy}}&=0,\\
\label{susybc2a5a}
D_{2}\varsigma^{\hat{l}}-[\Phi_{z},\Phi_{\overline{z}},\varsigma^{\hat{l}}]
\Bigl|_{\textrm{bdy}}&=0, 
&D_{z}\varsigma^{\hat{l}}
\Bigl|_{\textrm{bdy}}&=0, 
&D_{\overline{z}}\varsigma^{\hat{l}}
\Bigl|_{\textrm{bdy}}&=0,\\
\label{susybc2a5b}
D_{z}\varrho^{\hat{p}}
\Bigl|_{\textrm{bdy}}&=0, 
&D_{\overline{z}}\varrho^{\hat{p}}
\Bigl|_{\textrm{bdy}}&=0,\\
\label{susybc2a5c}
D_{2}\Phi_{z}+\frac12 [\varrho^{1},\varrho^{2},\Phi_{z}]
\Bigl|_{\textrm{bdy}}&=0, 
&D_{\overline{z}}\Phi_{z}
\Bigl|_{\textrm{bdy}}&=0,\\
\label{susybc2a5d}
D_{2}\Phi_{\overline{}}+\frac12
 [\varrho^{1},\varrho^{2},\Phi_{\overline{z}}]
\Bigl|_{\textrm{bdy}}&=0, 
&D_{z}\Phi_{\overline{z}}
\Bigl|_{\textrm{bdy}}&=0.
\end{align}
We see that 
the two bosonic scalars $\varphi^{\hat{i}}$ and $\varsigma^{\hat{l}}$ 
are also subject to the Dirichlet conditions in equation (\ref{susybc2a5}) 
and equation (\ref{susybc2a5a}) which are required by the other fivebrane. 
These conditions imply 
that they must be at fixed values 
so that they are the solutions to the Basu-Harvey type equations. 
Namely, the scalars $\varphi^{\hat{i}}$ and $\varsigma^{\hat{l}}$ 
obeying the boundary conditions (\ref{susybc2a5}) and (\ref{susybc2a5a}) 
have neither non-trivial solutions nor divergent behaviour 
as they are fixed at one end or at the other end.

As we already explained, 
the conditions $D_{\overline{z}}\Phi_{z}=0$ and 
$D_{z}\Phi_{\overline{z}}=0$ in 
equations (\ref{susybc2a5c}) and (\ref{susybc2a5d}) 
are the holomorphic and anti-holomorphic conditions 
rather than the Dirichlet boundary conditions. 
So they cannot completely fix the complex-valued 
$\Phi_{z}$ and $\Phi_{\overline{z}}$, 
which may still satisfy the Basu-Harvey like conditions 
in equations (\ref{susybc2a5c}) and (\ref{susybc2a5d}). 
Also note that the solutions to the Basu-Harvey-like equations 
of the complex-valued one-forms do not have divergent behaviour
as opposed to those of scalars with Nahm-like poles. 
Thus we expect that 
the bosonic degrees of freedom 
on the intersection of M2- and M5-branes inside the K3 
can be effectively described 
by means of the bosonic one-form $\Phi_{\alpha}$ by taking an
appropriate limit. 
%

%
%
Since the M5-brane and the M5'-brane 
break the isometry of the flat directions as 
\begin{align}
\label{m2m5m5dec1}
SO(6)_{345678}&\rightarrow SO(2)_{34}\times SO(4)_{5678}\nonumber\\
&\rightarrow SO(2)_{34}\times SO(2)_{56}\times SO(2)_{78}
\end{align}
via two projections, 
the 16 components of the fermionic fields 
in equation (\ref{twista2}) reduce to 
a pair of complex fermionic scalar fields 
in holomorphic and anti-holomorphic sectors, 
which we will call $\theta$ and $\overline{\theta}$,
and a pair of complex fermionic one-form fields 
in holomorphic and anti-holomorphic sectors, 
which we will call $p_{z}$ and $\overline{p}_{\overline{z}}$.

Here we want to pay special attention to 
the Basu-Harvey type supersymmetric boundary conditions in 
(\ref{susybc2a5}), (\ref{susybc2a5a}), (\ref{susybc2a5c}) 
and (\ref{susybc2a5d}) 
because they provide for us a hint about the effective theory 
as a topological sigma model. 
Given the Basu-Harvey type equations as the boundary conditions, 
the 3-bracket structure can survive at the boundary 
and the Hermitian 3-algebra can be constructed 
by a pair $(\mathfrak{so}(4),V)$ 
where $V$ is a faithful orthogonal representation 
of the Lie algebra $\mathfrak{so}(4)$ equipped with a 3-bracket. 
Quite interestingly, 
it was shown in \cite{deMedeiros:2008zh} that 
the Hermitian 3-algebra is generically embedded into 
a complex matrix Lie superalgebra $\mathfrak{sg}$ 
with an even subalgebra $\mathfrak{g}_{\mathbb{C}}$, 
the complexification of the corresponding Lie algebra $\mathfrak{g}$, 
and 
an odd subspace $V\oplus V^{*}$. 
In fact it has been pointed out more directly 
in \cite{Bielawski:2015wxa} that 
the Basu-Harvey type equations are sufficient conditions 
to realize the Lie superalgebra. 
In general the Jacobi identity of a Lie superalgebra 
consists of four components corresponding to the relationship 
between three elements of the Lie superalgebra; 
even-even-even, 
even-even-odd, 
even-odd-odd 
and odd-odd-odd. 
Among them the only non-trivial piece is the odd-odd-odd 
Jacobi identity and the Basu-Harvey type equation 
guarantees the odd-odd-odd Jacobi identity. 
Hence the appearance of the Basu-Harvey equations 
in the supersymmetric boundary conditions 
indicates that 
the target space of the effective sigma model
is a Lie superalgebra.
As we will explicitly see later, it is the Lie superalgebra
$\mathfrak{psl}(2|2)$ with even subalgebra 
$\mathfrak{sl}(2,\mathbb{C})\oplus \mathfrak{sl}(2,\mathbb{C})$.

\subsection{Gauge invariant boundary conditions}
%
%
So far we have determined the supersymmetric boundary conditions 
for the matter fields of the twisted BLG-model 
that would describe the M2-M5-M5' system (\ref{braneconf1a2}). 
However, 
we have not yet determined the boundary conditions for the gauge fields 
from supersymmetry 
because the supercurrent does not contain the field strength 
which demands the Neumann or the Dirichlet type boundary conditions 
for gauge fields. 
While there are many choices of boundary
conditions for gauge fields, 
we are especially interested in those which keep a full gauge symmetry. 
Boundary conditions for the gauge field in ABJM theory have previously been
studied in \cite{Berman:2009kj, Chu:2009ms, Berman:2009xd}
and in \cite{Chu:2009ms} a boundary action was introduced which preserved
the full gauge symmetry. 
However, here we will come up with an amazing result 
as a combination with the supersymmetric boundary conditions 
(\ref{susybc2a5})-(\ref{susybc2a5d}) on the twisted matter fields. 

Since the twisted Chern-Simons term (\ref{blg1a3}) 
whose variation produces a boundary term is not gauge invariant, 
we want to fix the boundary conditions 
for the gauge fields so that 
the gauge invariance of the bulk theory can be completely preserved 
\footnote{The boundary conditions which 
preserve only the diagonal part of the gauge symmetry group were
studied in \cite{Berman:2009xd}.}. 
First, let us consider the pure Chern-Simons action. 
The variation of the Chern-Simons action 
\begin{align}
\label{cs0a1}
S_{\mathrm{CS}}
&=\frac{k}{4\pi}\int_{M}d^{3}x \epsilon^{\mu\nu\lambda}
\mathrm{Tr}\left(
A_{\mu}\partial_{\nu}A_{\lambda}
+\frac{2i}{3}A_{\mu}A_{\nu}A_{\lambda}
\right)
\end{align}
yields 
\begin{align}
\label{cs0a2}
\delta S_{\textrm{CS}}
&=\frac{k}{4\pi}
\int_{M}d^{3}x \epsilon^{\mu\nu\lambda}
\mathrm{Tr}
\left(
\delta A_{\mu}F_{\nu\lambda}
\right)
+\frac{k}{4\pi}\int_{\partial M}
d^{2}x \epsilon^{\alpha\beta}\mathrm{Tr}\left(
\delta A_{\alpha}A_{\beta}
\right).
\end{align}
The second term does not automatically vanish on the boundary,
but boundary conditions which set one of the components of 
the gauge field to zero at the boundary 
can be chosen to make the Chern-Simons action invariant under the gauge transformation. 
The effect of such boundary conditions is that the Chern-Simons action can
be rearranged to show that the chosen component becomes a (bulk) Lagrange multiplier,
enforcing the constraint that the field strength in the orthogonal
directions vanishes \cite{Elitzur:1989nr}.
For a Lorentzian two-dimensional boundary  
one can choose the time-like $A_{0}|_{\textrm{bdy}}=0$, 
space-like $A_{1}|_{\textrm{bdy}}=0$, or light-like 
$A_{\pm}$ $:=$ $A_{0}\pm A_{1}|_{\textrm{bdy}}=0$ boundary conditions. 
The choice of boundary conditions determines the 
form of the boundary kinetic term. 
For example, the light-like boundary condition 
$A_{+}|_{\textrm{bdy}}=0$ leads to the constraint $F_{2-}=0$ 
\cite{Chu:2009ms}. 
The Euclidean two-dimensional boundary 
that we are now considering can be realized 
by performing the Wick rotation. 
The light-like boundary conditions become a holomorphic boundary condition 
\begin{align}
\label{gbc1a1}
A_{z}\Bigl|_{\textrm{bdy}}=\frac{1}{\sqrt{2}}
\left(
A_{0}-iA_{1}
\right)
\Bigl|_{\textrm{bdy}}=0 
\end{align}
and an anti-holomorphic boundary condition 
\begin{align}
\label{gbc1a2}
A_{\overline{z}}\Bigl|_{\textrm{bdy}}=\frac{1}{\sqrt{2}}
\left(
A_{0}+iA_{1}
\right)
\Bigl|_{\textrm{bdy}}=0, 
\end{align}
which respectively yield the conditions 
\begin{align}
\label{gbc1a3}
F_{2\overline{z}}=0
\end{align}
and 
\begin{align}
\label{gbc1a4}
F_{2z}=0.
\end{align}

Now let us first assume for simplicity that this
flatness condition can be solved as the pure gauge 
$A_{2}=g^{-1}\partial_{2}g$, 
$A_{z}=g^{-1}\partial_{z}g$ 
(or $A_{\overline{z}}=g^{-1}\partial_{\overline{z}}g$) 
where $g$ is a map from 
$\partial M = \Sigma$ at $x^2 = 0$ to the gauge group
$G$, and the map is arbitrarily smoothly extended to $x^2 > 0$. 
Substituting into the action we find 
\begin{align}
\label{cs0a3}
S_{\textrm{CS}}&=
\frac{k}{4\pi} \int_{M} d^{3}x 
\mathrm{Tr}
\left(
(g^{-1}\partial_{2}g)\partial_{z}(g^{-1}\partial_{\overline{z}}g)
-
(g^{-1}\partial_{\overline{z}}g)\partial_{z}(g^{-1}\partial_{2}g)
\right)\nonumber\\
&=
\frac{k}{4\pi} \int_{M} d^{3}x 
\mathrm{Tr}
\Biggl[
-(g^{-1}\partial_{2}g)(g^{-1}\partial_{z}g)(g^{-1}\partial_{\overline{z}}g)
+(g^{-1}\partial_{\overline{z}}g)(g^{-1}\partial_{z}g)(g^{-1}\partial_{2}g)\nonumber\\
&+
(g^{-1}\partial_{2}g)(g^{-1}\partial_{z}\partial_{-}g)
-(g^{-1}\partial_{\overline{z}}g)(g^{-1}\partial_{z}\partial_{2}g)
\Biggr].
\end{align}
Now integrate by parts with respect to $\overline{z}$ in the first 
and $x^{2}$ in the second. 
Then the integration by parts with respect to $x^{2}$ 
produces the standard kinetic term on the boundary 
while the first does not produce a boundary term. 
All other terms from the last line cancel between 
the two terms so that we are left with the WZW-model
\begin{align}
\label{cs1a1a}
S_{\textrm{WZW}}=
&-\frac{k}{8\pi}
\int_{\Sigma}
d^{2}x 
\mathrm{Tr}
\left(
g^{-1}\partial_{\alpha}g
\right)^{2}
-\frac{ik}{12\pi}
\int_{M}
d^{3}x \epsilon^{\mu\nu\lambda}\mathrm{Tr}
\left(
g^{-1}\partial_{\mu}g\cdot
g^{-1}\partial_{\nu}g\cdot
g^{-1}\partial_{\lambda}g
\right). 
\end{align}

Back to the case of the BLG-model with a boundary, 
let us define $A_{\mu}=A_{\mu 4i}^{(+)}\sigma_{i}$ 
and $\hat{A}_{\mu}=A_{\mu 4i}^{(-)}\sigma_{i}$ 
where $A^{(+)}_{\mu 4i}$ and $A^{(-)}_{\mu 4i}$ are 
self-dual and anti-self-dual parts of the gauge fields 
and the Pauli matrices $\sigma_{i}$ are normalized as 
$\mathrm{Tr}(\sigma_{i}\sigma_{j})=2\delta_{ij}$. 
Then the twisted Chern-Simons term (\ref{blg1a3}) in the BLG-model  
can be expressed as the $SU(2)_{k}$ $\times $ $SU(2)_{-k}$ 
quiver Chern-Simons term \cite{VanRaamsdonk:2008ft}
\begin{align}
\label{cs1a1}
S_{\textrm{CS}}&=
\frac{k}{4\pi}
\int_{M} d^{3}x \epsilon^{\mu\nu\lambda}
\left[
\mathrm{Tr}
\left(
A_{\mu}\partial_{\nu}A_{\lambda}
+\frac{2i}{3}A_{\mu}A_{\nu}A_{\lambda}
\right)
-
\mathrm{Tr}
\left(
\hat{A}_{\mu}\partial_{\nu}\hat{A}_{\lambda}
+\frac{2i}{3}\hat{A}_{\mu}\hat{A}_{\nu}\hat{A}_{\lambda}
\right)
\right]
\end{align}
with $k\in \mathbb{Z}$. 
Let us choose the boundary conditions 
\begin{align}
\label{bc1a4}
A_{z}\Big|_{\textrm{bdy}}&=0, &\hat{A}_{z}\Big|_{\textrm{bdy}}&=0,
\end{align} 
which require that 
\begin{align}
\label{bc1a5}
F_{2\overline{z}}&=0, &
\hat{F}_{2\overline{z}}&=0.
\end{align}
From the quiver Chern-Simons action (\ref{cs1a1}) 
we then find the boundary action 
\begin{align}
\label{cs1a2}
&S_{\textrm{$SU(2)_{k}$WZW}}[g]
+S_{\textrm{$SU(2)_{-k}$WZW}}[\hat{g}]\nonumber\\
=&
-\frac{k}{8\pi}
\int_{\Sigma}
d^{2}x 
\mathrm{Tr}
\left(
g^{-1}\partial_{\alpha}g
\right)^{2}
-\frac{ik}{12\pi}
\int_{M}
d^{3}x \epsilon^{\mu\nu\lambda}\mathrm{Tr}
\left(
g^{-1}\partial_{\mu}g\cdot
g^{-1}\partial_{\nu}g\cdot
g^{-1}\partial_{\lambda}g
\right)\nonumber\\
&
+\frac{k}{8\pi}
\int_{\Sigma}
d^{2}x 
\mathrm{Tr}
\left(
\hat{g}^{-1}\partial_{\alpha}\hat{g}
\right)^{2}
+\frac{ik}{12\pi}
\int_{M}
d^{3}x \epsilon^{\mu\nu\lambda}\mathrm{Tr}
\left(
\hat{g}^{-1}\partial_{\mu}\hat{g}\cdot
\hat{g}^{-1}\partial_{\nu}\hat{g}\cdot
\hat{g}^{-1}\partial_{\lambda}\hat{g}
\right).
\end{align}
In terms of $g$ and $\hat{g}$ 
the action (\ref{cs1a1}) now becomes 
a sum of the two WZW actions. 
As discussed in \cite{Elitzur:1989nr}, 
the measure is $\int [DA][D\hat{A}]\delta(F)\delta(\hat{F})$ 
$=\int [Dg][D\hat{g}]$ and there is no Jacobian in the change of
variables.

Now, in general the 
flat condition cannot necessarily be solved 
by a single-valued function on a curve 
as $g: \Sigma_{g}\times \mathbb{R}\rightarrow SU(2)$. 
The conjugacy class of a non-trivial holonomy 
of a flat connection around sources 
would lead to additional boundary degrees of freedom 
as the coadjoint action in the effective action
\cite{Moore:1989yh,Elitzur:1989nr}. 
However, even in the general case the WZW model would be part of the description,
along with a contribution from the Chern-Simons action involving the non-trivial
flat connections. As the moduli space of flat connections depends on the choice
of $\Sigma_g$ it may be more convenient to use an alternative method to
describe the Chern-Simons theory with a boundary. This involves adding new
boundary degrees of freedom, coupled to the bulk Chern-Simons action, in
such a way that gauge symmetry is preserved \cite{Witten:1991mm}. This approach
has been used in the context of the ABJM model \cite{Chu:2009ms, Berman:2009kj}.
In this approach it is clear that a WZW model will arise
from the bulk gauge field on any manifold with a boundary, even though
the full result including all ABJM matter fields and supersymmetry is not known
even in the simplest case of $M = \mathbb{R}^2 \times [0, L]$.
For our purposes the appearance of the WZW model is the key point, and at least
in the case of pure Chern-Simons theory the boundary conditions and boundary
degrees of freedom approaches are equivalent.

\section{Supergroup WZW Models}
\label{0sec4}
%

\subsection{\texorpdfstring{$PSL(2|2)$}{PSL(2|2)} WZW model and twisted BLG-model}

\subsubsection{Bosonic action}
%
Now we wish to collect the bosonic boundary conditions --
the supersymmetric boundary conditions
(\ref{susybc2a5})-(\ref{susybc2a5d}) 
and the gauge invariant boundary conditions (\ref{bc1a4})-(\ref{bc1a5}) --
to explore the effective boundary theory on the two membranes 
in the M2-M5-M5' system (\ref{braneconf1a2}). 

Let us introduce the complexified gauge fields
\begin{align}
\label{sgwzw0a0a}
{{\tilde{\mathcal{A}}_{z}\ }^{b}}_{a}
&={{\tilde{A}_{z}\ }^{b}}_{a}
+{f^{cdb}}_{a}\Phi_{\overline{z}c}\varphi^{*}_{d}
+{f^{cdb}}_{a}\Phi_{\overline{z}c}\varsigma^{*}_{d}\\
\label{sgwzw0a0b}
{{\tilde{\mathcal{A}}_{\overline{z}}\ }^{b}}_{a}
&={{\tilde{A}_{\overline{z}}\ }^{b}}_{a}
+{f^{cdb}}_{a}\Phi_{zc}\varphi_{d}
+{f^{cdb}}_{a}\Phi_{zc}\varsigma_{d}\\
\label{sgwzw0a0c}
{{\tilde{\mathcal{A}}_{2}\ }^{b}}_{a}
&={{\tilde{A}_{2}\ }^{b}}_{a}
+{f^{cdb}}_{a}\Phi_{\overline{z}c}\Phi_{zd}
+{f^{cdb}}_{a}\varphi_{c}\varsigma_{d}
+{f^{cdb}}_{a}\varrho_{c}\varrho^{*}_{d}
\end{align}
and the complexified field strength
\begin{align}
\label{sgwzw0a1a}
{\tilde{{\mathcal{F}}_{\mu\nu}}^{b} }_{a}
&=
\partial_{\nu}
{\tilde{{\mathcal{A}}_{\mu}}^{b} }_{a}-
\partial_{\mu}
{\tilde{{\mathcal{A}}_{\nu}}^{b} }_{a}
-{\tilde{{\mathcal{A}}_{\mu}}^{b} }_{c}{\tilde{{\mathcal{A}}_{\nu}}^{c} }_{a}
+{\tilde{{\mathcal{A}}_{\nu}}^{b} }_{c}{\tilde{{\mathcal{A}}_{\mu}}^{c} }_{a}
\end{align}
as well as the complexified scalars, e.g. 
$\varphi:=\frac{1}{\sqrt{2}}(\varphi^{1}-i\varphi^{2})$. 
Then by definition we have 
\begin{align}
\label{susybcc0a}
\tilde{\mathcal{F}}_{2\overline{z}}
&=\tilde{F}_{2\overline{z}}
-\left[
D_{2}\Phi_{z}+\frac12[\varrho,\varrho^{*},\Phi_{z}], 
\varphi,\  
\right]
-\left[
D_{2}\Phi_{z}+\frac12[\varrho,\varrho^{*},\Phi_{z}], 
\varsigma,\ 
\right]\nonumber\\
&
-\Bigl[\Phi_{z}, 
D_{2}\varphi-[\Phi_{\overline{z}},\Phi_{\overline{z}},\varphi]
,\ 
\Bigr]
-\Bigl[\Phi_{\overline{z}}, 
D_{2}\varsigma-[\Phi_{z},\Phi_{\overline{z}},\varsigma]
,\ 
\Bigr]\nonumber\\
&
+\Bigl[
D_{\overline{z}}\Phi_{\overline{z}},\Phi_{z},\ 
\Bigr]
+\Bigl[
\Phi_{\overline{z}}, 
D_{\overline{z}}\Phi_{z},\ 
\Bigr]
+\Bigl[ 
D_{\overline{z}}\varphi,
\varsigma, 
\ 
\Bigr]
+\Bigl[ 
\varphi, 
D_{\overline{z}}\varsigma, 
\ 
\Bigr]\nonumber\\
&+\Bigl[ 
D_{\overline{z}}\varrho,
\varrho^{*}, 
\ 
\Bigr]
+\Bigl[ 
\varrho,
D_{\overline{z}}\varrho^{*}, 
\ 
\Bigr].
\end{align}
Therefore both the supersymmetric boundary conditions
(\ref{susybc2a5})-(\ref{susybc2a5d}) 
and the gauge invariant boundary conditions (\ref{bc1a4})-(\ref{bc1a5}) 
can be unified as an equation
\begin{align}
\label{susybcc1a}
\tilde{\mathcal{F}}_{2\overline{z}}&=0
\end{align}
in terms of the complexified field strength (\ref{sgwzw0a1a}). 
%
%

It is remarkable that 
such a complexification of the gauge field and 
the simplification of the BPS equation are also encountered 
in the case of wrapped D3-branes 
on a holomorphic curve $\Sigma_{g}$ in K3 
(see e.g. \cite{Bershadsky:1995qy,Kapustin:2006pk}). 
In that case the effective theory can be described by 
the four-dimensional twisted $\mathcal{N}=4$ super Yang-Mills theory 
on $\Sigma_{g}$. 
A set of BPS equations on $\Sigma_{g}$ for $g>1$ is 
known to be Hitchin's equations; 
$F_{z\overline{z}}+i[\Phi_{z},\Phi_{\overline{z}}]=0$, 
$D_{z}\Phi_{\overline{z}}=0$ and $D_{\overline{z}}\Phi_{z}=0$. 
They can be summarized as the condition 
$\mathcal{F}_{z\overline{z}}=0$, which is the flatness condition 
on the complexified gauge field $\mathcal{A}_{z}:=A_{z}-i\Phi_{z}$. 
Moreover, the equation (\ref{susybcc1a}) reflects the fact that 
existence of high amounts of supersymmetry in Chern-Simons matter theories 
is inseparably bound up with gauge symmetry.

In trying to find the effective action 
of the boundary theory, 
we demand that it is classically scale invariant 
on the two-dimensional boundary $\Sigma_{g}$. 
To seek such a Lagrangian description, 
it is instructive to look at the bosonic action 
of the fully topologically twisted BLG-model 
on a general compact three-manifold $M$ 
\cite{Lee:2008cr}
\begin{align}
\label{sgwzw0a0d}
S_{\textrm{bosonic TBLG}}=\int_{M}d^{3}x 
& 
\left[
\frac{i}{2}\epsilon^{\mu\nu\lambda}
\left(
f^{abcd}\bm{\mathcal{A}}_{\mu ab}
\partial_{\nu}\bm{\mathcal{A}}_{\lambda cd}
+\frac23 {f^{cda}}_{g}f^{efgb}
\bm{\mathcal{A}}_{\mu ab}
\bm{\mathcal{A}}_{\nu cd}
\bm{\mathcal{A}}_{\lambda ef}
\right)
\right]
\nonumber\\
&+\sqrt{g}\mathrm{Tr}
\Biggl[
\frac12 
\left(
\bm{\mathcal{D}}^{\mu}\bm{\Phi}_{\mu}
-\frac{i}{3\sqrt{g}}\epsilon^{\mu\nu\lambda}
[\bm{\Phi}_{\mu},\bm{\Phi}_{\nu},\bm{\Phi}_{\lambda}]
\right)^{2}\nonumber\\
&+\frac14 
\left(
\bm{\mathcal{D}}_{\mu}\bm{\Phi}_{\nu}
-\bm{\mathcal{D}}_{\nu}\bm{\Phi}_{\mu}
\right)
\left(
\bm{\mathcal{D}}^{\mu}\bm{\Phi}^{\nu}
-\bm{\mathcal{D}}^{\nu}\bm{\Phi}^{\mu}
\right)\nonumber\\
&+\frac12 \bm{\mathcal{D}}^{\mu}\bm{\phi}^{I}
\bm{\mathcal{D}}_{\mu}\bm{\phi}^{I}
+\frac{1}{12}[\bm{\phi}^{I},\bm{\phi}^{J},\bm{\phi}^{K}]
[\bm{\phi}^{I},\bm{\phi}^{J},\bm{\phi}^{K}]\nonumber\\
&+\frac14 [\bm{\Phi}_{\mu},\bm{\phi}^{I},\bm{\phi}^{J}]
[\bm{\Phi}^{\mu},\bm{\phi}^{I},\bm{\phi}^{J}]
\Biggr]
\end{align}
where 
\begin{align}
\label{sgwzw0a0e}
\bm{\mathcal{A}}_{\mu ab}
&=A_{\mu ab}
-\frac{i}{2\sqrt{g}}\epsilon_{\mu\nu\lambda}{f^{cd}}_{ab}
\bm{\Phi}^{\nu}_{c}\bm{\Phi}_{d}^{\lambda},\\
\label{sgwzw0a0f}
(\bm{\mathcal{D}}_{\mu}X)_{a}
&=\partial_{\mu}X_{a}
-{\bm{\mathcal{A}}_{\mu}}^{b}\ _{a}X_{b}\nonumber\\
&=(D_{\mu} X)_{a} +\frac{i}{2\sqrt{g}}
\epsilon_{\mu\nu\lambda}
[\bm{\Phi}^{\nu},\bm{\Phi}^{\lambda},X]_{a}
\end{align}
are the three-dimensional versions of complexified objects 
while $\bm{\Phi}_{\mu}$ and $\bm{\phi}^{I}$ are component fields of 
the bosonic $SO(3)$ one-form field 
and the five bosonic scalar fields respectively. 

However, for the partially twisted BLG-model on $\Sigma_{g}$ 
together with boundaries, 
the complexification comes about in a slightly different way 
from (\ref{sgwzw0a0e}) and (\ref{sgwzw0a0f}) 
according to the breakdown of the rotational symmetry; 
$SO(3)\rightarrow SO(2)$. 
This allows the complexified scalar fields 
to enter the complexification as in 
(\ref{sgwzw0a0a})-(\ref{sgwzw0a0c}). 
Note that the $x^{2}$ component of the complexified gauge fields, 
$\mathcal{A}_{2}$ is now identified with a bosonic scalar field 
and contains additional contributions from complexified scalar fields 
in our definition (\ref{sgwzw0a0c}).  
Furthermore the partially twisted action takes a different form 
in terms of 
the modified complexified gauge fields (\ref{sgwzw0a0a})-(\ref{sgwzw0a0c}). 
In the fully twisted action (\ref{sgwzw0a0d}) there are four types of 
classically scale invariant terms on a Riemann surface $\Sigma_{g}$ 
which can contribute to the effective boundary action; 
\begin{enumerate}
\renewcommand{\labelenumi}{(\roman{enumi})}
\item the twisted Chern-Simons term of the complexified gauge fields 
in the first line, 
\item the quadratic term 
$(D_{2}\Phi_{\alpha}-D_{\alpha}\phi)^{2}$ 
with $\phi$ being the bosonic scalar fields in the third line,
\item  the kinetic terms $(D_{\alpha}\phi)^{2}$ of the bosonic scalar fields 
in the fourth line, 
\item the potential terms of the form 
$[\Phi_{\alpha},\phi,\phi]^{2}$ 
in the fifth line. 
\end{enumerate}

Now we point out that under the supersymmetric boundary conditions 
(\ref{susybc2a5})-(\ref{susybc2a5d}) which 
are encoded by the complexified gauge fields 
(\ref{sgwzw0a0a})-(\ref{sgwzw0a0c}) 
as equation (\ref{susybcc1a}),  
all the possible terms (i)-(iv) can be formally collected 
as the twisted Chern-Simons term
\begin{align}
\label{sgwzw0a1}
S_{\textrm{bosonic TBLG}}
&=\int_{\Sigma_{g}\times I} d^{3}x 
\left[
\frac12 \epsilon^{\mu\nu\lambda}
\left(
f^{abcd}\mathcal{A}_{\mu ab}\partial_{\nu}\mathcal{A}_{\lambda cd}
+\frac23 {f^{cda}}_{g}f^{efgb}
\mathcal{A}_{\mu ab}\mathcal{A}_{\nu cd}\mathcal{A}_{\lambda ef}
\right)
\right]
\end{align}
of the complexified gauge fields 
(\ref{sgwzw0a0a})-(\ref{sgwzw0a0c}). 
This implements the complexification of 
the twisted Chern-Simons term (\ref{blg1a3}).
One can therefore view the supersymmetric boundary conditions 
(\ref{susybc2a5})-(\ref{susybc2a5d}) 
and the gauge invariant boundary conditions 
(\ref{bc1a4})-(\ref{bc1a5}) 
as the complexified gauge invariant boundary condition 
(\ref{susybcc1a}) in the twisted Chern-Simons term (\ref{sgwzw0a1}). 
Following the previous logic, 
we can now get the bosonic boundary action. 

The twisted Chern-Simons term (\ref{blg1a3}) can be rewritten as 
a sum of two Chern-Simons actions as in (\ref{cs1a1}).  
With the aid of the gauge invariant boundary conditions 
(\ref{bc1a4})
they give rise to a sum of two WZW actions (\ref{cs1a2}),
although as previously discussed we cannot exclude additional contributions
from non-trivial flat connections.
Thus the twisted Chern-Simons term (\ref{sgwzw0a1}) 
of the complexified gauge fields $\mathcal{A}_{\mu}$ 
with the boundary condition (\ref{susybcc1a}) 
generates the boundary action as 
a sum of two $SL(2,\mathbb{C})$ WZW actions 
\begin{align}
\label{sgwzw00}
S_{\textrm{bosonic}}
=&S_{\textrm{$SL(2,\mathbb{C})_{k}$WZW}}[g]
+S_{\textrm{$SL(2,\mathbb{C})_{-k}$WZW}}[\hat{g}]\nonumber\\
=&
-\frac{k}{8\pi}
\int_{\Sigma_{g}}
d^{2}x 
\mathrm{Tr}
\left(
g^{-1}\partial_{\alpha}g
\right)^{2}
-\frac{ik}{12\pi}
\int_{M}
d^{3}x \epsilon^{\mu\nu\lambda}\mathrm{Tr}
\left(
g^{-1}\partial_{\mu}g\cdot
g^{-1}\partial_{\nu}g\cdot
g^{-1}\partial_{\lambda}g
\right)\nonumber\\
&
+\frac{k}{8\pi}
\int_{\Sigma_{g}}
d^{2}x 
\mathrm{Tr}
\left(
\hat{g}^{-1}\partial_{\alpha}\hat{g}
\right)^{2}
+\frac{ik}{12\pi}
\int_{M}
d^{3}x \epsilon^{\mu\nu\lambda}\mathrm{Tr}
\left(
\hat{g}^{-1}\partial_{\mu}\hat{g}\cdot
\hat{g}^{-1}\partial_{\nu}\hat{g}\cdot
\hat{g}^{-1}\partial_{\lambda}\hat{g}
\right).
\end{align}

\subsubsection{Including fermionic terms}
As discussed at the end of section~\ref{secSUSYbc}, on general grounds we
expect to have a supergroup structure. Obviously the natural expectation is
that including the fermionic fields will enhance the $SL(2) \times SL(2)$
WZW model to a $PSL(2|2)$ WZW model. We will first review the form of the
$PSL(2|2)$ WZW action \cite{Bhaseen:1999nm,Gotz:2006qp,Bershadsky:1999hk} and then discuss how this can arise from the
twisted Chern-Simons theory with our fermionic field content.

Let us begin by considering the $SL(2|2)$ WZW-model
\begin{align}
\label{sgwzw01}
S_{SL(2|2)_{k}}[s]=
-\frac{k}{8\pi}\int_{\Sigma_{g}}
d^{2}x \mathrm{Str}
\left(s^{-1}\partial_{\alpha}s\right)^{2}
-\frac{ik}{12\pi}\int_{M}d^{3}x \epsilon^{\mu\nu\lambda}
\mathrm{Str}\left(
s^{-1}\partial_{\mu}s\cdot 
s^{-1}\partial_{\nu}s\cdot
s^{-1}\partial_{\lambda}s
\right)
\end{align}
for supermatrices
\begin{align}
\label{sgwz01b}
s&=\left(
\begin{array}{cc}
A&B\\
C&D\\
\end{array}
\right) \in SL(2|2)
\end{align}
with $A$ and $D$ being bosonic matrix elements of
$SL(2,\mathbb{C})$ and $B$ and $C$ being fermionic matrix elements.
Here a supertrace $\mathrm{Str}$ is defined as
$\mathrm{Str}(s)=\mathrm{Tr}(A)-\mathrm{Tr}(D)$.
The supergroup element $s\in SL(2|2)$
admits the Gauss decomposition \cite{Isidro:1993er}
\begin{align}
\label{sgwz01c}
s&=
\exp(u)
\left(
\begin{array}{cc}
\mathbb{I}&0\\
\overline{\theta}&\mathbb{I}\\
\end{array}
\right)
\left(
\begin{array}{cc}
g&0\\
0&\hat{g}\\
\end{array}
\right)
\left(
\begin{array}{cc}
\mathbb{I}&\theta\\
0&\mathbb{I}\\
\end{array}
\right)\nonumber\\
&=\exp(u)\left(
\begin{array}{cc}
g&g \theta\\
\overline{\theta} g&\overline{\theta} g\theta+\hat{g}\\
\end{array}
\right)
\end{align}
with $u\in \mathbb{C}$ and
$g, \hat{g}$ $\in SL(2,\mathbb{C})$.
The action (\ref{sgwzw01}) satisfies
the Polyakov-Wiegmann identity \cite{Polyakov:1984et}
\footnote{Providing one replaces trace with a supertrace,
the Polyakov-Wiegmann identity for supergroups
takes the same form as that for ordinary groups
according to the cyclic property of a supertrace.}
\begin{align}
\label{pwid1}
S[s_{1}s_{2}]&=S[s_{1}]+S[s_{2}]
+\frac{k}{2\pi} \int d^{2}x \mathrm{Str}
\left(
s^{-1}_{1}\partial_{\overline{z}}s_{1}\partial_{z}s_{2}s_{2}^{-1}
\right).
\end{align}

Now, $SL(2|2)$ has a normal $U(1)$ subgroup
consisting of multiples of the identity.
As discussed e.g.\ in
\cite{Bershadsky:1999hk} the $SL(2|2)$-invariant metric is degenerate. However,
treating the U(1) symmetry as a gauge symmetry and quotienting by this U(1)
results in a $PSL(2|2)$ WZW model, with a non-degenerate invariant metric. Since
$PSL(2|2)$ has bosonic subgroup $SL(2) \times SL(2)$ this also gives the
minimal embedding of the bosonic $SL(2) \times SL(2)$ WZW model into a
supergroup WZW model.

Although $PSL(2|2)$ has no representation of supermatrices,
one can descend to $PSL(2|2)$ from $SL(2|2)$
by identifying supermatrices $s\in SL(2|2)$
which differ by a scalar multiple.
Using the Polyakov-Wiegmann identity (\ref{pwid1})
we can show that
\begin{align}
\label{pwid2}
S_{SL(2|2)_{k}}[e^{u}s]
&=S_{SL(2|2)}[e^{u}]+S_{SL(2|2)_{k}}[s]
+\frac{k}{2\pi}\int_{\Sigma_{g}} d^{2}x 
\partial_{\overline{z}}u 
\mathrm{Str}(s\partial_{z} s^{-1})\nonumber\\
&=S_{SL(2|2)_{k}}[s].
\end{align}
This states that the action (\ref{sgwzw01}) is invariant
after multiplying the supermatrices $s\in SL(2|2)$
with a scalar factor $\exp(u)$.
In other words,
the $PSL(2|2)$ WZW action is equivalent to the $SL(2|2)$ WZW action.

Applying the Polyakov-Wiegmann identity (\ref{pwid1})
to the decomposition (\ref{sgwz01c})
one can rewrite the $PSL(2|2)$ WZW model (\ref{sgwzw01}) as
\begin{align}
\label{sgwz01d0}
S_{PSL(2|2)_{k}}[s]&=
S_{PSL(2|2)_{k}}
\left[
\left(
\begin{array}{cc}
\mathbb{I}&0\\
\overline{\theta}&\mathbb{I}\\
\end{array}
\right)
\right]
+
S_{PSL(2|2)_{k}}
\left[
\left(
\begin{array}{cc}
g&0\\
0&\hat{g}\\
\end{array}
\right)
\right]
+
S_{PSL(2|2)_{k}}
\left[
\left(
\begin{array}{cc}
\mathbb{I}&\theta\\
0&\mathbb{I}\\
\end{array}
\right)
\right]
\nonumber\\
&+\frac{k}{2\pi}
\int_{\Sigma_{g}}d^{2}x \mathrm{Str}
\left(
\begin{array}{cc}
0 & g^{-1}\partial_{\overline{z}}g\partial_{z}\theta\\
\partial_{\overline{z}}\overline{\theta} (\partial_{z} g) g^{-1}
& \hat{g}^{-1}\partial_{\overline{z}}\overline{\theta} g\partial_{z}\theta\\
\end{array}
\right).
\end{align}
The first and third terms vanish
because contributions to supertraces can arise only from
non-trivial bosonic submatrices.
Then the final result is
\begin{align}
\label{sgwz01d}
S_{PSL(2|2)_{k}}[s]&=
S_{SL(2,\mathbb{C})_{k} \textrm{WZW}}[g]
+S_{SL(2,\mathbb{C})_{-k} \textrm{WZW}}[\hat{g}]\nonumber\\
&-\frac{k}{2\pi}
\int_{\Sigma_{g}}d^{2}x \mathrm{Tr}
\left(
\hat{g}^{-1}
\partial_{\overline{z}}\overline{\theta} 
g
\partial_{z}\theta 
\right).
\end{align}
The first two terms are the sum of two $SL(2,\mathbb{C})$ WZW models
which we have encountered in the bosonic boundary action in
(\ref{sgwzw00}).
Notice that the opposite level comes from
the definition of the supertrace.

Let us now proceed to discuss how this supergroup WZW model can arise from
the fermionic degrees of freedom we have.
Recall that the supersymmetric boundary conditions 
in the topologically twisted BLG model allow for 
the spin-zero fermionic fields $\theta$, $\overline{\theta}$ 
and spin-one fermionic fields $p_{z}$, $\overline{p}_{\overline{z}}$. 
We identify $\theta$, $\overline{\theta}$ with the fermionic fields in the
supergroup WZW model. There could be a field redefinition in this relation,
but this would just correspond to a different parametrisation of the supergroup
elements. However, we note that the supergroup action does not include the
fields $p_{z}$, $\overline{p}_{\overline{z}}$.

In constructing the boundary action with fermionic terms, we again demand that
the possible boundary terms are scale invariant
at the classical level.
In two dimensions, the spin-zero and spin-one fermionic fields have scaling dimensions zero and one respectively. 
Without the couplings of the fermions to the bosons, 
one can write a conformally invariant action \cite{Friedan:1985ge}
\begin{align}
\label{fer1a0}
S&=\frac{1}{\pi}
\int d^{2}x 
\left(
p_{z}\partial_{\overline{z}}\overline{\theta}
+\overline{p}_{\overline{z}}\partial_{z}\theta
\right).
\end{align}
This is the fermionic ghost system 
with central charge $c=-2$, 
the so-called symplectic fermions \cite{Kausch:2000fu}. 
However, we should consider other possible terms 
which stem from the terms in the twisted BLG theory, 
i.e.\ the fermionic kinetic term 
$\left(\overline{\Psi},\Gamma^{\mu}D_{\mu}\Psi\right)$ 
and the interaction term 
$\left(\overline{\Psi},\Gamma^{IJ}[X^{I},X^{J},\Psi]\right)$. 
This means that 
the boundary terms are quadratic in fermionic fields, with up to one derivative acting on the fermions,
and that 
they can also contain the bosonic matrix fields $g$, $\hat{g}$ 
and their inverses $g^{-1}$, $\hat{g}^{-1}$ of $SL(2,\mathbb{C})$. 
Taking into account the scale invariance, we could have the following 
possible boundary terms including the fermionic fields;
$\theta$, $\overline{\theta}$, $p_{z}$, $\overline{p}_{\overline{z}}$, 
and the bosonic fields; $g$, $\hat{g}$, $g^{-1}$, $\hat{g}^{-1}$:
\begin{enumerate}
\renewcommand{\labelenumi}{(\roman{enumi})}
\item terms involving two fermionic scalar fields 
and no fermionic one-form field
\begin{align}
\label{fer1a1}
\partial_{z}\theta \hat{g}^{-1}
\partial_{\overline{z}}\overline{\theta}g, 
\ \ \ \ \ 
\partial_{z}\theta
 \hat{g}\partial_{\overline{z}}\overline{\theta}g^{-1},
\end{align}
\item terms involving a fermionic scalar field 
and a fermionic one-form field
\begin{align}
\label{fer1a2}
p_{z}\hat{g}^{-1}\partial_{\overline{z}}\overline{\theta}g, 
\ \ \ \ \ 
\overline{p}_{\overline{z}}\hat{g}\partial_{z}\theta g^{-1},
\end{align}
\item terms involving no fermionic scalar field and two fermionic
      one-form fields
\begin{align}
\label{fer1a3}
p_{z}\hat{g}^{-1}\overline{p}_{\overline{z}}g,
\ \ \ \ \  
p_{z}\hat{g}\overline{p}_{\overline{z}}g^{-1}.
\end{align}
\end{enumerate}

Now, the terms in (\ref{fer1a1}) have two derivatives and do not obviously arise from
the fermionic terms in the BLG theory. However,
we note that 
the fermionic one-form fields $p_{z}$, $\overline{p}_{\overline{z}}$ 
have no kinetic terms and they therefore 
should be treated as auxiliary fields. 
After integrating them out we are expected to be left 
with the terms as in (\ref{fer1a1}) 
which only contain the fermionic scalar fields $\theta$,
$\overline{\theta}$. 
Thus the 
fermionic boundary degrees of freedom
can be encoded in the interaction term
\begin{align}
\label{fer1a4}
S_{\textrm{int}} 
&=\int_{\Sigma_{g}}
d^{2}x \mathrm{Tr}
\left(
\kappa_{A}
\partial_{z}\theta \hat{g}^{-1}\partial_{\overline{z}}\overline{\theta}g
+\kappa_{B}
\partial_{z}\theta \hat{g}\partial_{\overline{z}}\overline{\theta}g^{-1}
\right)
\end{align}
where $\kappa_{A}$ and $\kappa_{B}$ are constants. 

Note that we didn't directly derive the fermionic action from the
twisted BLG action. In fact, since
$g$ and $\hat{g}$ are dimensionless, it is consistent with dimensional analysis
that there could be a variety of terms with
$(\partial_z \theta) (\partial_{\overline{z}}\overline{\theta})$ in the
final fermionic boundary action. Likewise the precise form of the coupling to
$g$ and $\hat{g}$ may seem somewhat arbitrary. However, we would expect the
requirement of
conformal invariance (at the quantum level) to be highly restrictive.
As we have seen above, including the above fermionic terms with
$\kappa_{A}=-\frac{k}{2\pi}$, $\kappa_{B}=0$ gives the $PSL(2|2)$ WZW model, so
this is certainly consistent with all our requirements for the action.

One plausible argument to constrain the allowed fermionic terms is to note that
the bosonic WZW action has the obvious global symmetry
$$ g \rightarrow u g v^{-1} \; , \;\;\; \hat{g} \rightarrow \hat{u} \hat{g} \hat{v}^{-1} . $$
If we assume the classical action has this symmetry when we include the
fermions, we must assign specific transformation properties to the fermions or
they could not couple to the bosonic fields. As the bulk fermions were in
the bifundamental representation, we likewise expect the boundary fermions to
allow coupling of $g$ to $\hat{g}$. In order for this to be possible we take the following transformation rules:
\begin{eqnarray}
\theta \rightarrow v \theta \hat{v}^{-1} & , &
	\overline{\theta} \rightarrow \hat{u} \overline{\theta} u^{-1} , \\
p_z \rightarrow v p_z \hat{v}^{-1} & , &
	\overline{p}_{\overline{z}} \rightarrow \hat{u} \overline{p}_{\overline{z}} u^{-1} .
\end{eqnarray}
Similar transformation rules are possible with different choices of
$u \leftrightarrow v$ or $\hat{u} \leftrightarrow \hat{v}$ but these just differ
by field redefinitions, by multiplying the fermions on the left or right by
$g$, $\hat{g}$ or their inverses. With this particular convention we see that
only the first term in (\ref{fer1a3}) is allowed while both terms in
(\ref{fer1a2}) are possible. If these three terms are all present in the
action\footnote{If any of these terms has zero coefficient, the fermionic part
of the action will vanish after integrating out $p_z$ and/or
$\overline{p}_{\overline{z}}$. So, we would be left with only the bosonic
$SL(2) \times SL(2)$ WZW model.}
then integrating out $p_z$ will produce exactly (\ref{fer1a4}) with
$\kappa_B = 0$ and the non-zero value of $\kappa_A$ will just correspond to the
normalisation of $\theta$ and $\overline{\theta}$. Even assuming the global
symmetry, this argument is not quite complete as there are possible terms
similar to those in (\ref{fer1a2}) where the derivative acts on $g$ or
$\hat{g}$. Allowing all such terms will generate several additional terms after
integrating out $p_z$, e.g.\ $g^{-1}\partial g \theta \hat{g}^{-1} \overline{\partial} \overline{\theta}$. However, as far as we are aware, such possible
fermionic couplings do not give rise to a CFT, so the requirement of
conformal invariance will not allow such terms.

While we have not given a rigorous derivation, we believe that this is the unique result arising from
the topologically twisted BLG-model on $\Sigma_{g}$
by choosing the supersymmetric and gauge invariant boundary conditions.
Indeed, from the form of the bosonic part of the action, and the requirement of
a supergroup structure, this is essentially the only possibility (other than adding additional fields which do not arise from the bulk theory.) If we simply
demand conformal invariance, it is possible that other fermionic interactions
are allowed\footnote{These would not respect the global symmetry present in
the classical action of the bosonic sector, but we cannot directly rule out such a possibility.}.
However, we are not aware of any such models with the same bosonic
action (and where the fermions are coupled to the bosonic fields.) We do note
that there is the interesting possibility discussed in \cite{Bershadsky:1999hk}
that there is a family of $PSL(N|N)$ WZW models which are conformal even when
the coefficients of the kinetic term and the WZW term are independent. We have
only discussed the case with the standard relation between these coefficients
as that is what we expect to get from the Chern-Simons theory. However, it
would be interesting to understand what role, if any, such deformations have
in this context.

To summarise, we have obtained the $PSL(2|2)$ WZW model from 
the topologically twisted BLG-model on $\Sigma_{g}$ 
by choosing the supersymmetric and gauge invariant boundary conditions. 
In view of the form of the action (\ref{sgwz01d}), 
we see that 
a heuristic construction of highly supersymmetric conformally invariant 
gauge theories in three dimensions as the quiver Chern-Simons matter theories   
with opposite integer levels 
is intimately related to the structure of the supergroup WZW actions 
underlying the supertrace. 
Due to the wrong sign of the kinetic term, 
we do not expect this to be a unitary theory or even to directly arise from one
by an analytic continuation. 
So one could not extract the dynamical properties 
of the M2-branes. 
However, the theory we are now considering 
is the effective field theory 
on the Euclidean $\Sigma_{g}$ wrapped by the M2-branes. 
The resulting theory, which is different from the 
original physical theory via topological twisting, 
could only capture the topological properties, 
or the BPS spectrum of the curved M2-branes 
wrapping a holomorphic Riemann surface $\Sigma_{g}$ 
as a topological field theory.
We expect that 
it can play a similar role as other proposed effective theories 
arising from curved world-volumes of branes, 
e.g., 
two-dimensional topological sigma models
for wrapped D3-branes on Riemann surfaces in K3 
\cite{Bershadsky:1995qy}, 
$SL(2,\mathbb{C})$ Chern-Simons theory 
for wrapped M5-branes on 3-manifolds
in Calabi-Yau three-folds \cite{Dimofte:2011ju}, 
Vafa-Witten theory 
for wrapped M5-branes on 4-manifolds
in $G_{2}$ manifolds \cite{Gadde:2013sca}. 
Indeed the supergroup WZW-model is known 
to be a topological sigma model \cite{Isidro:1993er,Bershadsky:1999hk} 
and also has been used to compute the Alexander polynomials $\Delta$ 
\cite{Rozansky:1992rx}. 
Remarkably it has been proven in \cite{MR1650308} that 
any A-polynomials which occur as the Alexander polynomials 
can occur as the Seiberg-Witten invariant of an 
irreducible homotopy K3. 
We expect to be able to address these relations with our physical setup. 
In particular, 
for the effective theory of the topological M-strings 
in the brane configuration (\ref{braneconf1a2}), 
the level would be $k=1$ 
since only it can realize two flat M2-branes in a flat space. 
Therefore we propose the $PSL(2|2)$ WZW model with level $k=1$ 
as the effective action of the two topological M-strings.

\subsection{\texorpdfstring{$GL(N|N)$}{GL(N|N)} WZW model and twisted ABJM-model}
Let us generalize our discussion to the case 
of an arbitrary number of coincident M2-branes 
in the brane configuration (\ref{braneconf1a2}). 
The ABJM-model \cite{Aharony:2008ug}, 
which is a three-dimensional $\mathcal{N}=6$ superconformal 
$U(N)_{k}\times U(N)_{-k}$ Chern-Simons matter theory   
has been proposed as the low-energy world-volume effective theory 
of $N$ M2-branes probing $\mathbb{C}^{4}/\mathbb{Z}_{k}$. 
The theory involves four complex scalar fields $Y^{A}$, 
four Weyl spinor fields $\psi_{A}$ 
and two types of gauge fields $A_{\mu}, \hat{A}_{\mu}$. 
The theory has $SU(4)_{R}$ R-symmetry group 
as well as $U(1)_{B}$ flavor symmetry group. 
$Y^{A}$ and $\psi_{A}$ are the matter fields transforming 
as the $(\bm{N},\overline{\bm{N}})$ bi-fundamental representation 
of the $U(N)_{k}\times U(N)_{-k}$ gauge group with 
$U(1)_{B}$ charge $+1$, 
while $Y_{A}^{\dag}$ and $\psi^{\dag A}$ are those transforming 
as the $(\overline{\bm{N}},\bm{N})$ anti-bi-fundamental representation with 
$U(1)_{B}$ charge $-1$. 
The upper and lower indices $A,B,\cdots=1,2,3,4$ correspond to the
$\bm{4}$ and $\overline{\bm{4}}$ of the $SU(4)_{R}$ R-symmetry 
and baryonic charges $1$ and $-1$ respectively, while
$A_{\mu}$ are the $U(N)$ Chern-Simons gauge fields
of level $+k$ 
and $\hat{A}_{\mu}$ are the $U(N)$ Chern-Simons gauge fields 
of level $-k$. 
The gauge fields transform as the trivial representation 
of $SU(4)_{R}$ $\times$ $U(1)_{B}$.

If we try to get the low-energy effective theory 
of $N$ topological M-strings 
by carrying out the topological twist on the ABJM-model, 
it is necessary to consider the effect of the $U(1)_{B}$ charge. 
The global symmetry $SU(4)_{R}\times U(1)_{B}$ has 16 currents. 
However, when $k=1,2$ the monopole operators 
provide us with 12 symmetry generators 
so that the global symmetry is enhanced to $SO(8)_{R}$ 
with 28 generators. 
Thus the $\mathcal{N}=6$ supersymmetry of the ABJM-model is 
expected to be enhanced to $\mathcal{N}=8$ for $k=1$ and $k=2$ 
by taking into account the baryon symmetry $U(1)_{B}$ 
\cite{Aharony:2008ug,Gustavsson:2009pm,Kwon:2009ar}. 
As discussed in \cite{Labastida:1997rg},  
a topological twisting procedure generically can be regarded as 
a gauging of an internal symmetry group 
by adding to the original action 
the coupling of the internal current to the spin connection 
and one can also take such an internal symmetry as a baryon symmetry. 

Now we attempt to twist the ABJM-model 
by first decomposing the R-symmetry as 
\begin{align}
\label{abjmtwist0a}
SU(4)_{R}\rightarrow SU(3)_{R}\times U(1)_{R} .
\end{align} 
Then we define a generator $\mathfrak{s}'$ of the $SO(2)_{E}'$ as
\begin{align}
\label{abjmtwista1}
\mathfrak{s}'&=\mathfrak{s}-\frac12 T_{R} -\frac12 T_{B}
\end{align}
where $\mathfrak{s}$ is a generator of the original rotational group $SO(2)_{E}$, 
$T_{R}$ is a generator of $U(1)_{R}$ and $T_{B}$ is a generator of $U(1)_{B}$. 
The branching of the representation for the decomposition 
$SU(4)_{R}$ $\rightarrow$ 
$SU(3)_{R}\times U(1)_{R}$ 
is
\begin{align}
\label{sbjmtwist0b}
\bm{6}&\rightarrow \bm{3}_{2}\oplus \overline{\bm{3}}_{-2}\nonumber\\
\bm{4}&\rightarrow \bm{3}_{-}\oplus \bm{1}_{3}\nonumber\\
\overline{\bm{4}}&\rightarrow \overline{\bm{3}}_{+}\oplus \bm{1}_{-3}.
\end{align}
The twisting 
$SO(2)_{E}\times SU(4)_{R}\times U(1)_{B}$ $\rightarrow$ 
$SO(2)_{E}'\times SU(3)_{R}$ 
reduces the supersymmetry parameter $\omega$ as follows:
\begin{align}
\label{abjmtwista2a}
\omega&:\ \ \ \bm{6}_{+}\oplus \bm{6}_{-}
\rightarrow \bm{3}_{0}\oplus \bm{3}_{2}
\oplus \overline{\bm{3}}_{-2}\oplus \overline{\bm{3}}_{0}.
\end{align}
The appearance of the six covariantly constant spinors 
indicates that the twisting procedure corresponds to 
the M-theory background (\ref{braneconf1a1}) 
since K3 breaks half of the supersymmetry. 
After the twisting 
$SO(2)_{E}\times SU(4)_{R}\times U(1)_{B}$ $\rightarrow$ 
$SO(2)_{E}'\times SU(3)_{R}$, the fields transform as
\begin{align}
\label{abjmtwista2b}
Y^{A}&:\ \ \ \bm{4}_{0}
\rightarrow \bm{3}_{0}\oplus \bm{1}_{-2}\nonumber\\
Y_{A}^{\dag}&:\ \ \ \overline{\bm{4}}_{0}
\rightarrow \overline{\bm{3}}_{0}\oplus \bm{1}_{2}\nonumber\\
\psi_{A}&:\ \ \ \overline{\bm{4}}_{+}\oplus \overline{\bm{4}}_{-}
\rightarrow \overline{\bm{3}}_{0}\oplus \bm{1}_{2}\oplus \overline{\bm{3}}_{-2}\oplus \bm{1}_{0}\nonumber\\
\psi^{\dag A}&:\ \ \ \bm{4}_{+}\oplus \bm{4}_{-}
\rightarrow \bm{3}_{2}\oplus \bm{1}_{0}\oplus \bm{3}_{0}\oplus \bm{1}_{-2}.
\end{align}

We see that the twisted ABJM-model comprises 
scalar fields $\bm{3}_{0}\oplus \overline{\bm{3}}_{0}$ with six components, 
bosonic one forms $\bm{1}_{2}\oplus \bm{1}_{-2}$ giving two components, 
fermionic scalar fields $\bm{3}_{0}\oplus \bm{1}_{0}$ 
$\oplus \overline{\bm{3}}_{0}\oplus \bm{1}_{0}$ with eight components 
and another eight components from fermionic one-form fields $\bm{1}_{2}\oplus \bm{3}_{2}$ 
$\oplus$ $\bm{1}_{-2}\oplus \overline{\bm{3}}_{-2}$.
In other words, the twisted ABJM theory 
has exactly the same number of bosonic and fermionic field components 
as (\ref{twist1a1}) and (\ref{twista2}) in the twisted BLG theory. 
By imposing the appropriate supersymmetric boundary conditions 
on these fields, 
we find holomorphic, anti-holomorphic fermionic scalars 
$\theta$, $\overline{\theta}$ 
as well as fermionic one-form fields 
$p_{z}$, $\overline{p}_{\overline{z}}$. 
Therefore we are led to regard the above twisted theories 
as the source of the low-energy effective description of 
$N$ topological M-strings. 

Since the twisting requires gauging 
both the $U(1)_{R}$ and $U(1)_{B}$ symmetries, 
a straightforward decomposition of gamma matrices and spinors cannot work. 
However, we would like to make a few remarks on the effective theory. 
First, 
the $U(N)\times U(N)$ Chern-Simons action should produce a sum of the two WZW
actions with the holomorphic boundary conditions 
$A_{z}\bigl|_{\textrm{bdy}}=0$, $\hat{A}_{z}\bigl|_{\textrm{bdy}}=0$ 
as in (\ref{bc1a4}) and (\ref{bc1a5}) 
since the topological twisting does not affect 
the gauge fields and the Chern-Simons action. 
Second, 
the ABJM-model is shown to be written 
in terms of the 3-algebra \cite{Bagger:2008se}, 
which enables us to define complexified gauge fields 
as in (\ref{sgwzw0a0a})-(\ref{sgwzw0a0c}). 
This would promote the $U(N)\times U(N)$ gauge fields 
to the complexified $GL(N)\times GL(N)$ gauge fields. 
Third, 
the ABJM-model has the BPS boundary conditions for
the bosonic scalar fields analogous to the Basu-Harvey equations 
which may represent $N$ M2-branes ending on the M5-brane 
\cite{Terashima:2008sy}. 
It has been also argued in \cite{Bielawski:2015wxa} that 
these are sufficient conditions for the presence of 
the Lie superalgebra 
$\mathfrak{gl}(N|N)$
with even subalgebra 
$\mathfrak{gl}(N,\mathbb{C})$ $\oplus$ $\mathfrak{gl}(N,\mathbb{C})$. 
Finally, the symplectic fermions which are necessary to obtain 
the free field realization of the supergroup WZW models and the associated 
affine Lie superalgebra wonderfully and automatically appear in the field content (\ref{abjmtwista2b}) 
of the topologically twisted ABJM theory. 
Given the remarks above, 
the topologically twisted ABJM-model on $\Sigma_{g}$ 
with the supersymmetric and gauge invariant 
boundary conditions would provide the $GL(N|N)$ WZW action 
\footnote{Instead of supertrace 
we have introduced the non-degenerate bilinear form 
$\langle\cdot,\cdot\rangle$ 
for the non-semisimple $\mathfrak{gl}(N|N)$.}
\begin{align}
\label{sgwz02a} 
S_{GL(N|N)_{k}}[s]=
&
-\frac{k}{8\pi}\int_{\Sigma_{g}}
d^{2}x 
\left
\langle 
s^{-1}\partial_{\alpha}s,
s^{-1}\partial^{\alpha}s
\right
\rangle \nonumber\\
&
-\frac{ik}{24\pi}\int_{M}d^{3}x \epsilon^{\mu\nu\lambda}
\left
\langle 
s^{-1}\partial_{\mu}s, 
[s^{-1}\partial_{\nu}s,s^{-1}\partial_{\lambda}s]
\right
\rangle.
\end{align}

We note that while the supermatrix $s\in GL(N|N)$ may have 
the Gauss decomposition \cite{Isidro:1993er}
\begin{align}
\label{sgwz02b}
s&=
\left(
\begin{array}{cc}
\mathbb{I}&0\\
\overline{\theta}&\mathbb{I}\\
\end{array}
\right)
\left(
\begin{array}{cc}
g&0\\
0&\hat{g}\\
\end{array}
\right)
\left(
\begin{array}{cc}
\mathbb{I}&\theta\\
0&\mathbb{I}\\
\end{array}
\right)
\end{align}
with $g,\hat{g}$ $\in$ $GL(N)$ being 
Grassmann-even matrix elements 
and $\theta,\overline{\theta}$ being 
Grassmann-odd matrix elements, 
the Polyakov-Wiegmann relation may be generalized for 
the bilinear form $\langle \cdot,\cdot\rangle$. 
If we consider 
the effective theories of the $N$ topological M-strings 
in the brane system (\ref{braneconf1a2}), 
they can be realized for the level $k=1$ associated 
with a flat background geometry.  
The $GL(N|N)$ WZW models 
have previously been proposed as an explicit realization 
of topological conformal field theories \cite{Isidro:1993er}.  
From the free field realization (\ref{sgwz02a}) 
upon the Gauss decomposition (\ref{sgwz02b}) 
the theory has been argued to be represented as 
the superposition of two decoupled parts with 
$SL(N,\mathbb{C})$ and $U(1)$ symmetries, 
both of which constitute topological conformal field theories.

More generally we can consider 
other twisted $\mathcal{N}=6$ Chern-Simons matter theories.  
To preserve $\mathcal{N}=6$ supersymmetry 
the gauge groups of Chern-Simons matter theories are not arbitrary 
and the other allowed options are 
$U(N)\times U(M)$ and $SO(2)\times Sp(N)$
\cite{Aharony:2008gk,Hosomichi:2008jb}. 
These $\mathcal{N}=6$ superconformal Chern-Simons theories 
can be also formulated in terms of the Lie 3-algebra 
by relaxing the conditions on the triple product 
so that it is not real and antisymmetric in all three indices 
\cite{Bagger:2008se}. 
Evidently it is straightforward to extend our discussion 
to these $\mathcal{N}=6$ Chern-Simons matter theories 
by following the same argument as the ABJM-model 
although the M-theory interpretation is much less transparent. 
Consequently we would obtain the WZW models of the supergroups 
$GL(N|M)$ and $OSp(2|N)$ from 
the $\mathcal{N}=6$ 
$U(N)\times U(M)$ and $SO(2)\times Sp(N)$ 
Chern-Simons matter theories 
by performing partial topological twists on $\Sigma_{g}$ 
and imposing the supersymmetric and gauge invariant boundary 
conditions. 
It would also be interesting to analyse cases with $\mathcal{N}=5$ or
$\mathcal{N}=4$ supersymmetry where again the gauge group must be the
even part of a supergroup \cite{Gaiotto:2008sd,Hosomichi:2008jd,Hosomichi:2008jb,Bergshoeff:2008bh,deMedeiros:2008zh,deMedeiros:2009eq}.

\section{Discussion}
\label{0sec5}
The present work should be extended 
in a number of directions. 
From the field theory point of view, 
we propose the novel correspondence between 
the supergroup WZW models and 
the topologically twisted Chern-Simons matter theories. 
This would give a way to resolve the puzzle that 
the well-known correspondence 
between WZW and Chern-Simons theories for ordinary compact groups 
\cite{Witten:1988hf,Elitzur:1989nr,Witten:1991mm,Axelrod:1989xt} 
is not available for generic supergroups 
\cite{BarNatan:1991rn,Rozansky:1992zt,Mikhaylov:2014aoa,Mikhaylov:2015nsa}. 
It is known that 
the $GL(N|N)$ WZW models 
are topological field theories of cohomological type 
as they have $c=0$ and their stress-energy tensors are BRST exact. 
An issue worthy of investigation is the interpretation of these topological
theories in their own right. 
In \cite{Rozansky:1992td} 
the multivariable Alexander-Conway knot polynomial 
\cite{MR1501429,MR0258014} of links in $S^{3}$ 
has been explicitly obtained 
from the $S$ and $T$ matrices of the $GL(1|1)$ WZW model. 
Also the $GL(N|N)$ WZW models have been 
expected to produce the Alexander-Conway polynomial 
\cite{Rozansky:1992rx,Isidro:1993er}. 
It is rather interesting to note that 
in \cite{MR1650308} the homotopy K3 surfaces \cite{MR0262545}
have been constructed 
from knots in three-manifolds,
and the Seiberg-Witten invariants of these manifolds 
have been shown to be given by the Alexander polynomials of the knots. 
We expect that our M-theory framework 
will prove useful to understand and generalize the problem 
in that 
the M2-branes wrapped on a two-cycle in K3 are described 
by the supergroup WZW models having a conjectural relation 
to the Alexander knot polynomials. 
This is currently under investigation.

Our proposed 3d-2d relation -- i.e.\ the relation between 
three-dimensional supersymmetric Chern-Simons matter theories,
realized on the worldvolume of M2-branes,
and two-dimensional supergroup WZW models --
has an analogue in one higher dimension. 
In that case the relation is between 
four-dimensional $\mathcal{N} = 4$ super Yang-Mills theories,
realized on the worldvolume of D3-branes,
and three-dimensional Chern-Simons theories.
In particular, the rich structure 
of boundary conditions for $\mathcal{N} = 4$ 
super Yang-Mills theories was explored in \cite{Gaiotto:2008sa},  
describing D3-branes ending on
various types of 5-branes in type IIB string theory. 
There are indications that 
the boundary theories admit the Lie superalgebraic structure 
\cite{Gaiotto:2008sd,Kapustin:2009cd}. 
In fact, Mikhaylov and Witten \cite{Mikhaylov:2014aoa} 
established that the defect theories of 
the topologically twisted $\mathcal{N}=4$ super Yang-Mills theories 
can be described by supergroup Chern-Simons theories. 
They can be embedded in type IIB string theory 
having the D3-branes ending on both sides of an NS5-brane.
E.g.\ the simplest case with supergroup 
$U(m\vert n)$ arises when $m$ D3-branes end on one side,  
and $n$ D3-branes on the other side, of an NS5-brane. 

In this article we have only described in detail 
very specific M2-M5 configurations. 
There are several possible generalizations 
and we hope to report on other cases in future work.
An obvious question is what happens 
when M2-branes end from both sides of an M5-brane. 
In the string theory case bifundamental matter arises from open
strings connecting the D3-branes on either side. 
It is not obvious what the
corresponding feature is in M-theory, 
but we expect it is possible to derive
the low energy theory from the Chern-Simons theories 
and boundary conditions.
Other natural generalizations are 
to relax the boundary conditions on the scalar fields 
by taking orientations of the M5-branes other than the M5-M5' case we
investigated, or by allowing a finite separation between the M5-branes. 
Such generalizations will have additional scalar degrees of 
freedom and not result in purely topological theories. 
However, our expectation is that 
the supergroup WZW models will continue to describe the 
internal degrees of freedom of the M2-brane boundary while additional 
fields will describe the transverse degrees of freedom of the M2-brane 
boundaries within the M5-branes. 

A related approach to describing theories with boundaries is to add boundary
degrees of freedom rather than imposing (some) boundary conditions. 
In this context Belyaev and van Nieuwenhuizen \cite{Belyaev:2008xk}
studied the boundary degrees of freedom required to preserve half the bulk
supersymmetry.
This approach was applied to the ABJM theory 
with a boundary in \cite{Berman:2009kj} 
resulting in a partial description of the boundary theory 
which was sufficient to derive the boundary scalar potential 
for certain amounts of preserved 
supersymmetry, while more general boundary conditions were analyzed in
\cite{Berman:2009xd}. 
In \cite{Chu:2009ms} the same systems were analyzed 
with particular focus on the boundary degrees of freedom required 
to preserve the Chern-Simons gauge symmetry. 
Some aspects of supersymmetry were considered, 
but the fully supersymmetric M-string theory was not derived. 
In light of the various M2-M5 configurations
described above we hope to develop the boundary action approach to derive the
full (half of original bulk) supersymmetric 
and fully gauge invariant boundary action. 
Work on this is currently in progress and 
we hope to report results in the near future. 
Applied to the special configuration considered in this article, 
such an approach would give an independent derivation 
of the emergence of the supergroup. 

Finally, we note that there is recent work
\cite{Armoni:2015jsa} on coupled Chern-Simons--WZW systems 
with less supersymmetry 
arising from D3-D5 configurations, 
and a theory of this type has been proposed \cite{Niarchos:2015lla}
which should flow at low energies 
to the $\mathcal{N} = (4,2)$ or $\mathcal{N} = (4,4)$ M-string theories. 
In part the results in this article, 
and our anticipated results for the more general cases, 
are naturally a higher supersymmetric analogue, 
and for M-strings would be a direct derivation 
of the low energy theory. 
However, it is not clear at this stage 
whether there is any way to directly study 
such a theory by flowing from the theory 
proposed in \cite{Niarchos:2015lla}.

\subsection*{Acknowledgements}
We are grateful to 
Chong-Sun Chu, Chun-Chung Hsieh, Kazuo Hosomichi and Seiji Terashima 
for useful comments and discussions. 
We especially thank Mir Faizal 
for collaboration at an early stage of this work 
and Roger Bielawski 
for helpful explanation of his work. 
Research of T.O. is supported by 
MOST under the Grant No.104-2811-056. 
D.J.S. is supported in
part by the STFC Consolidated Grant ST/L000407/1.

\bibliographystyle{utphys}
\bibliography{ref}

\end{document}